\documentclass[twocolumn]{aastex6}
\usepackage{graphicx}
\usepackage{epsfig}
\usepackage{natbib}
\usepackage{multirow}
\usepackage{amsmath}

\newcommand{\eg}{{\rm e.g.\ }}
\newcommand{\cf}{{\rm cf.\ }}
\newcommand{\etal}{{\rm et al.\thinspace}}

\newcommand{\km}{{\rm\thinspace km}}

\newcommand{\s}{{\rm\thinspace s}}

\newcommand{\kmps}{\hbox{$\km\s^{-1}\,$}}

\newcommand{\fesc}{$f_{\rm esc}$}

\newcommand{\lya}{Ly$\alpha$}
\newcommand{\lyg}{Ly$\gamma$}

\newcommand{\hi}{H\thinspace{\sc i}}
\newcommand{\hii}{H\thinspace{\sc ii}}
\newcommand{\nii}{[N\thinspace{\sc ii}]}

\newcommand{\mgii}{Mg\thinspace{\sc ii}}
\newcommand{\oiii}{[O\thinspace{\sc iii}]}
\newcommand{\oiiib}{O\thinspace{\sc iii}]}
\newcommand{\oii}{[O\thinspace{\sc ii}]}

\newcommand{\neiii}{[Ne\thinspace{\sc iii}]}

\newcommand{\heii}{He\thinspace{\sc ii}}
\newcommand{\hei}{He\thinspace{\sc i}}

\newcommand{\cii}{C\thinspace{\sc ii}}
\newcommand{\ciip}{C\thinspace{\sc ii}]}
\newcommand{\ciii}{C\thinspace{\sc iii}]}
\newcommand{\fciii}{[C\thinspace{\sc iii}]}
\newcommand{\civ}{C\thinspace{\sc iv}}

\newcommand{\Msol}{\hbox{\thinspace M$_{\sun}$}}
\newcommand{\Zsol}{\hbox{\thinspace Z$_{\sun}$}}

\slugcomment{Draft of \today}

\shorttitle{\ciii\ in Emission Line Galaxies}

\begin{document}

\title{Photoionization Models for the Semi-Forbidden \ciii~$\lambda$1909 Emission in Star-Forming Galaxies}

\author{A. E. Jaskot, \altaffilmark{1} \&\
	S. Ravindranath \altaffilmark{2}}
\altaffiltext{1}{Department of Astronomy, Smith College, Northampton, MA 01063, USA.}
\altaffiltext{2}{Space Telescope Science Institute, Baltimore, MD 21218, USA.}

\begin{abstract}

The increasing neutrality of the intergalactic medium at $z>6$ suppresses \lya\ emission, and spectroscopic confirmation of galaxy redshifts requires detecting alternative UV lines. The strong \fciii~$\lambda$1907$+$\ciii~$\lambda$1909 doublet frequently observed in low-metallicity, actively star-forming galaxies is a promising emission feature. We present CLOUDY photoionization model predictions for \ciii\ equivalent widths (EWs) and line ratios as a function of starburst age, metallicity, and ionization parameter. Our models include a range of C/O abundances, dust content, and gas density. We also examine the effects of varying the nebular geometry and optical depth. Only the stellar models that incorporate binary interaction effects reproduce the highest observed \ciii\ EWs. The spectral energy distributions from the binary stellar population models also generate observable \ciii\ over a longer timescale relative to single-star models. We show that diagnostics using \ciii\ and nebular \heii~$\lambda$1640 can separate star-forming regions from shock-ionized gas. We also find that density-bounded systems should exhibit weaker \ciii\ EWs at a given ionization parameter, and \ciii\ EWs could therefore select candidate Lyman continuum-leaking systems. In almost all models, \ciii\ is the next strongest line at $< 2700$ \AA\ after \lya, and \ciii\ reaches detectable levels for a wide range of conditions at low metallicity. \ciii\ may therefore serve as an important diagnostic for characterizing galaxies at $z>6$.  

\end{abstract}

\section{Introduction}
\label{sec:intro}

Nebular emission lines are among the most valuable diagnostics of galaxy properties. By tracing gas photoionized by massive stars, emission lines can measure a galaxy's star formation rate \citep[\eg][]{kennicutt83, kennicutt98rev}. Furthermore, line ratios of species with different ionization potentials reveal the spectral energy distribution (SED) of the incident radiation \citep[\eg][]{kaler78, baldwin81, stasinska96, rigby04, zastrow13b}. These lines provide a key observational test of stellar SED models and can distinguish between stellar and non-stellar ionizing sources. The observed strengths of nebular emission lines also depend strongly on the properties of the interstellar medium, such as its average electron density, electron temperature, composition, and dust extinction \citep[\eg][]{searle72, osterbrock89, calzetti94, kewley02}. Historically, most nebular diagnostics were predominantly optical emission lines. Since these common diagnostics fall in the infrared at higher redshift ($z > 7$), beyond the wavelengths accessible to most large telescopes, many recent studies have explored the diagnostic potential of shorter wavelength emission lines for star-forming galaxies \citep[\eg][]{erb10, levesque14, feltre16}.

One line of particular interest is the semiforbidden \ciii~$\lambda$1909 transition, which is often blended with the nearby forbidden \fciii~$\lambda$1907 line. Strong \ciii\ doublet emission appears in spectra of low-metallicity galaxies at both low and high redshift \citep{shapley03, erb10, leitherer11, bayliss14, stark14, stark15, stark16, rigby15}. Among low-redshift starbursts, lower metallicity galaxies tend to have higher \ciii\ equivalent widths (EWs), presumably due to higher electron temperatures in low-metallicity gas \citep{heckman98}. With a sample of 70 star-forming regions from galaxies at $z=0-4$, \citet{rigby15} show that strong \ciii\ emission with EW$>5$\AA\ only occurs below a metallicity of $12+{\rm log(O/H)}=8.4$ or $\sim$0.5 \Zsol. High \ciii\ EWs appear to be common in low-mass ($M_* \lesssim 10^9 $\Msol) galaxies at $z>1.5$. Of the 17 low-mass galaxies at $z=1.5-3$ presented in \citet{stark14}, 76\%\ have \ciii\ EWs greater than 4\AA. Recent observations probing \ciii\ in low-mass galaxies at even higher redshift likewise find high \ciii\ EWs: $\sim$10\AA\ for a lensed galaxy at $z=3.6$ \citep{bayliss14} and $\sim$22.5\AA\ in two galaxies at $z=6.0-7.7$ \citep{stark15, stark16}.

Strong \ciii\ emission from low-metallicity galaxies has a number of potential uses. In their sample of 17 $z=1.5-3$ galaxies, \citet{stark14} find that after \lya, \ciii\ is always the next strongest ultraviolet (UV) line. They therefore propose that \ciii\ emission could provide spectroscopic confirmation of galaxy redshifts, particularly at $z>6$, when the high neutral fraction of the intergalactic medium impedes \lya\ propagation. In combination with other UV lines, such as \heii~$\lambda$1640, \ciii\ could also distinguish between photoionization by active galactic nuclei (AGN), photoionization from young stars, and shock ionization \citep{villarmartin97, feltre16}.

Nevertheless, the origin of strong \ciii\ emission is not fully understood. In particular, some low-metallicity galaxies, such as I Zw 18, and more massive galaxies at $z\sim2$, lack strong \ciii\ emission \citep{rigby15}. Galaxies may need both low metallicity and high ionization to produce high EW \ciii\ \citep[\eg][]{erb10, stark14}. However, as \citet{rigby15} point out, I Zw 18 should fulfill both criteria, although its \ciii\ is weak. Furthermore, the dependence of \ciii\ strength on metallicity is not simple. On the one hand, higher electron temperatures in low-metallicity galaxies will increase collisional excitations to the upper levels of the \ciii\ transitions. On the other hand, the C/O abundance ratio appears to decrease with declining metallicity \citep[\eg][]{garnett04, erb10}. This metallicity dependence of C/O may either arise from the weaker winds in low-metallicity massive stars or from the longer timescale necessary for lower mass stars to enrich interstellar gas in carbon \citep{henry00, chiappini03, akerman04, erb10}. Different C/O ratios among galaxies would also lead to variation in their observed \ciii\ emission.

Observationally, high EW \ciii\ emission appears in galaxies with intense star formation and strong nebular emission lines. At low redshift, the star-forming galaxies with the strongest \ciii\ emission have Wolf-Rayet features indicative of a recent starburst \citep{rigby15}. These same galaxies show extreme optical nebular emission line equivalent widths \citep{rigby15}, similar to the optical line strengths of extreme emission line galaxies at $z=1-8$ \citep[\eg][]{vanderwel11, labbe13, smit14}. The SEDs of high-redshift \ciii-emitting galaxies likewise suggest strong optical emission lines, with  \oiii~$\lambda$5007 and H$\beta$ EWs of several hundred {\AA}ngstroms \citep{stark14}. \lya\ emission may also correlate with \ciii\ emission, albeit with substantial scatter \citep{shapley03, rigby15}. High-redshift \lya-emitting galaxies are often low-mass, highly star-forming systems \citep[\eg][]{gawiser07, hagen14, finkelstein15}, and strong \lya\ emission appears more common at high redshift \citep[\eg][]{stark11, zheng14}. 

These galaxy samples suggest that extreme emission line galaxies and \lya\ emitters (LAEs) have properties conducive to \ciii\ emission line formation. At low redshift, the rare ``Green Pea" (GP) galaxies exhibit both extreme optical nebular emission and strong \lya, which argues that they may be the ideal low-redshift sample to examine \ciii\ production. Discovered in the Sloan Digital Sky Survey, the GPs are compact, low-mass, starburst galaxies with low ($\sim 0.2$ \Zsol) metallicities \citep{amorin10, izotov11} and \oiii~$\lambda$5007 EWs as high as 2000 \AA\ \citep{cardamone09}. They share many similarities with $z>2$ LAES including compact morphologies, high specific star formation rates, low dust extinction, and nebular line ratios that indicate high ionization parameters \citep[\eg][]{cardamone09, amorin12, jaskot13, nakajima14, malhotra12, hagen14, gawiser07}. Indeed, the GPs typically exhibit strong \lya\ emission \citep{jaskot14, henry15}. The GPs may also be analogs of cosmic reionizers; their optical line ratios, \lya\ profiles, and weak low-ionization absorption suggest Lyman continuum (LyC) escape \citep{jaskot13, jaskot14, verhamme15}. \citet{izotov16, izotov16b} recently confirmed LyC escape from five low-redshift GPs, the highest LyC escape fractions yet measured at low redshift. The GPs' similarities to high-redshift galaxies suggest that they may serve as a valuable test of models of UV line production for the low-metallicity, highly star-forming galaxies that populate the high-redshift universe.

In this paper, we present CLOUDY photoionization model grids \citep{ferland98} that reveal the dependence of \ciii~$\lambda$1909 emission strength on stellar population age, ionization parameter, metallicity, and other nebular parameters. We assess the causes of strong \ciii\ emission and compare the model grids with existing \ciii\ detections. We examine the relationship between \ciii\ line strength and other optical and UV lines and use the observed line ratios of the GPs to predict the \ciii\ emission strength in extreme emission line galaxies. Finally, we address the implications of these predictions for future studies of galaxies at $z>6$.

\section{Photoionization Modeling}
\label{sec:models}
We use the CLOUDY photoionization code \citep{ferland98}, version 13.1, to explore predicted UV line strengths and \fciii~$\lambda$1907$+$\ciii~$\lambda$1909 (hereafter \ciii~$\lambda$1909) as a function of different ionizing SEDs and nebular parameters. We adopt a solar metallicity of $12+{\rm log(O/H)}=8.69$ and \Zsol$=0.014$ \citep{asplund09} throughout our analysis.

\subsection{Input SEDs}
\label{sec:models:seds}
For the incident radiation field for our modeled nebulae, we consider different stellar population models, star formation histories, and stellar metallicities. Spectral synthesis models of massive star populations vary greatly in their predicted ionizing SEDs \citep[\eg][]{leitherer14, wofford16, stanway16}. Here, we test predictions from three different spectral synthesis codes: (1) the Padova models for single, non-rotating stars, with asymptotic giant branch stars included, as implemented in the Starburst99 code \citep{leitherer99, leitherer14}, (2) the Geneva models for single, rotating stars as implemented in Starburst99 \citep{georgy13, leitherer14}, and (3) the Binary Population and Spectral Synthesis (BPASS) models, version 2.0 (\citealt{stanway16}; Eldridge \etal\ in prep.), which include the effects of both stellar rotation and binary interactions. 

For each stellar population model, we adopt a broken power law initial mass function (IMF) with a slope of $-1.3$ from $0.1-0.5$ \Msol\ and a slope of $-2.35$ from $0.5-100$ \Msol. The high-mass slope corresponds to the \citet{salpeter55} IMF. We consider two simple star formation histories for comparison with other studies: a single-age stellar population that formed from an instantaneous burst and a composite stellar population that formed from continuous star formation at a constant rate. We explore a range of ages for each of these scenarios. 

Each stellar population model differs in the available stellar metallicity options. The Padova models have evolutionary tracks for stars of four metallicities from $Z=$0.0004 to 0.02, the rotating Geneva models have tracks for five metallicities between $Z=0.001-0.040$, and the BPASS models have tracks for 11 metallicities between $Z=0.001-0.040$. Since \ciii\ emission is absent in high metallicity galaxies \citep{rigby15}, we focus on the stellar population models where $Z\leq0.014$.

\begin{figure*}
\plotone{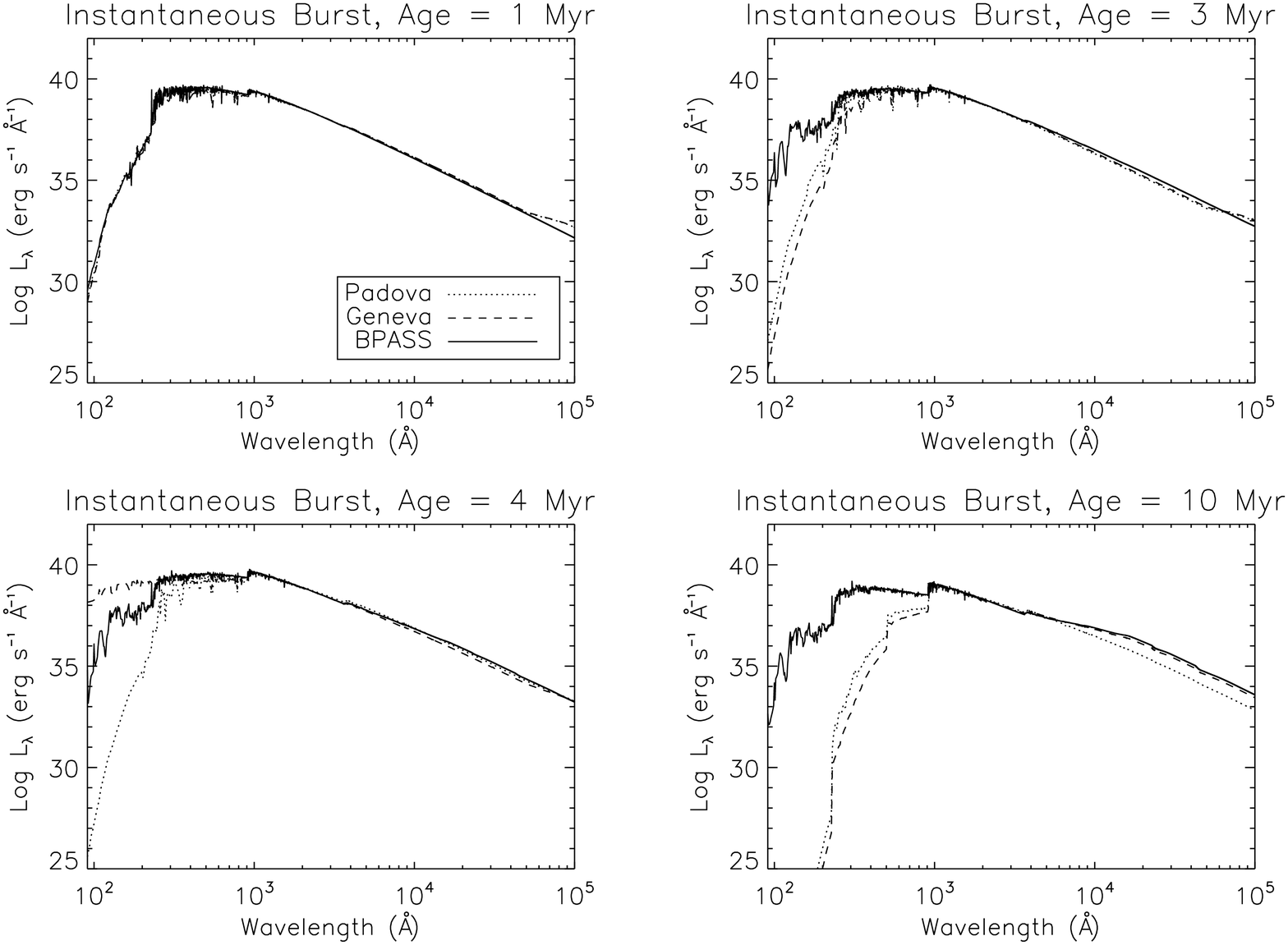}
\caption{A comparison of the SEDs from the Padova single-star evolutionary tracks at $Z=0.004$, Geneva single-star tracks with rotation at $Z=0.002$, and BPASS binary stellar evolution tracks at $Z=0.002$. The panels show the predicted stellar SEDs at ages of 1, 3, 4, and 10 Myr after an instantaneous burst of star formation. All SEDs are normalized to the same luminosity at 1500 \AA.}
\label{fig:sedcompare}
\end{figure*}

Figure~\ref{fig:sedcompare} compares the ionizing SEDs from the three spectral synthesis codes at different ages after an instantaneous burst of star formation. The SEDs correspond to a metallicity of $Z=0.002$ for the Geneva and BPASS models and $Z=0.004$ for the Padova models; the Padova models do not include SEDs for $Z=0.002$. At the youngest ages, all three models give similar predictions, although the BPASS models emit a slightly higher flux at the shortest UV wavelengths. However, the effect of binary interactions becomes increasingly important at later times. After 3 Myr, stronger mass loss rates, mass transfer, and binary mergers increase the UV flux from the binary models relative to the single star models \citep{stanway16}. Although the Padova models show a brief uptick in UV flux at 4 Myr from a strong Wolf-Rayet stellar population, the BPASS models exhibit the strongest UV fluxes at both earlier and later ages.

\subsection{Nebular Parameters}
\label{sec:models:neb}
To calculate the nebular emission, we run each input SED model through the CLOUDY photoionization code. We vary the adopted nebular geometry, ionization parameter, hydrogen density, and atomic and dust abundances as described below. The predicted nebular emission corresponds to the emission from gas surrounding a single star-forming region (see \citealt{charlot01} and \citealt{gutkin16} for an alternative approach). In reality, the observed emission from galaxies will be a luminosity-weighted average of the emission from multiple star-forming regions, each of which may have different stellar and nebular properties. The simple models presented here approximate individual starburst regions, galaxies with emission dominated by a single luminous region, or the emission from multiple regions with similar properties. 

\subsubsection{Geometry}
\label{sec:models:geometry}
Following \citet{stasinska15}, we consider two different nebular geometries: a filled sphere and a thin shell of gas. We define the inner nebular radius of the nebula, $R_{\rm in}$, in terms of the Str{\"o}mgren radius, $R_S$, where $R_{\rm in}=f_S R_S$. For the filled sphere models, $f_S=0.03$, and for the thin shell models, $f_S=3$. 

We also consider the effect of optical depth by truncating the nebular gas at different outer radii. We parameterize the optical depth by comparing the cumulative H$\beta$ flux produced inside each outer radius. We define $f_{{\rm H}\beta} = F_{{\rm H}\beta} / F_{{\rm H}\beta, {\rm RB}}$, where $F_{{\rm H}\beta, {\rm RB}}$ is the total integrated H$\beta$ flux for the optically thick, radiation-bounded nebula. Optically thin, density-bounded nebulae have $f_{{\rm H}\beta} < 1$, and smaller values of $f_{{\rm H}\beta}$ correspond to increasingly truncated nebulae and higher escape fractions of ionizing radiation.

\subsubsection{Ionization Parameter}
\label{sec:models:u}
The ionization parameter, $U$, compares the density of ionizing photons incident on a cloud to the nebula's electron density. At a given radius, the dimensionless ionization parameter
\begin{equation}
U=\frac{Q_{\rm H}}{4 \pi r^2 c n_{\rm H}},
\label{eqn:u}
\end{equation}
where $Q_{\rm H}$ is the number of ionizing photons emitted per second, $n_{\rm H}$ is the hydrogen number density, and $c$, the speed of light, makes $U$ dimensionless. Studies from the literature differ in their treatment of $U$; some studies evaluate $U$ at the Str{\"o}mgren radius, $R_S$ \citep[\eg][]{stark14}, while others evaluate $U$ at the inner nebular radius, $R_{in}$ \citep[\eg][]{erb10}.
Here, we follow \citet{charlot01} and \citet{stasinska15} and calculate the volume-averaged ionization parameter, $\bar{U}$:
\begin{equation}
\bar{U}=(\frac{3}{4 \pi})^{1/3}\frac{\alpha_{\rm B}^{2/3}}{c}(Q_{\rm H} n_{\rm H} \epsilon^2)^{1/3}((1+f_S^3)^{1/3}-f_S),
\end{equation}
where $\alpha_{\rm B}$ is the case B recombination coefficient and $\epsilon$ is the volume filling factor of the gas. We adopt $\alpha_{\rm B} = 2.6 \times 10^{-13}$ cm$^3$ s$^{-1}$, the value for 10$^4$ K gas from \citep{storey95}. For comparison to values in the literature, we relate $\bar{U}$ to $U(R_S)$ and $U(R_{\rm in})$:
\begin{equation}
\bar{U}=3 U(R_S) ((1+f_S^3)^{1/3}-f_S)
\end{equation}
and
\begin{equation}
\bar{U}=3 U(R_{\rm in}) f_S^2 ((1+f_S^3)^{1/3}-f_S).
\end{equation}
We run CLOUDY models for ${\rm log}(n_{\rm H}) =$ 1, 2, 3, and 4 cm$^{-3}$ and log($\bar{U})=$ -4, -3, -2, and -1 with each geometry (sphere and shell). Varying $\epsilon$ has no apparent effects on our results, and we therefore only consider models with $\epsilon=1$. While most galaxies may never reach ionization parameters as high as log $U =-1$ \citep[\eg][]{yeh12}, the nebular emission of some extreme blue compact dwarf galaxies, such as the GPs, could be consistent with log $U\geq-1.5$ \citep{stasinska15}. In these galaxies, young starburst regions may dominate the nebular emission, leading to the high inferred ionization parameters.

\subsubsection{Abundances}
\label{sec:models:Z}

For each photoionization model, the gas-phase oxygen metallicities match the stellar metallicities of the input ionizing SED. We adopt the CLOUDY ``Interstellar Medium (ISM)" abundances for Ne, S, Ar, and Cl and scale them by the same factor as the O abundance. For the He abundance, we interpolate the abundances given in \citet{mcgaugh91}, assuming a solar metallicity of $12+{\rm log(O/H)}=8.69$ \citep{asplund09}.

Both massive and intermediate-mass stars produce C and N, and C/O and N/O ratios vary among galaxies \citep[\eg][]{edmunds78, garnett04}. We therefore consider models with a range of C/O and N/O ratios. Following \citet{erb10} and \citet{stark14}, we scale the default C/O and N/O ratios from the CLOUDY ISM abundance set by factors of 0.05, 0.15, 0.25, 0.45, 0.65, 0.85, and 1. The resulting C/O ratios are 0.04, 0.12, 0.20, 0.35, 0.51, 0.67, and 0.79, and the N/O ratios are 0.01, 0.04, 0.06, 0.11, 0.16, 0.21, and 0.25. We apply the same scale factor for both C and N. 

We vary the dust abundance by adopting dust-to-metal ratios of $\xi = $ 0.1, 0.3, and 0.5 \citep[\eg][]{stark14}. In addition, we adjust the abundances of some heavy elements to account for depletion onto dust grains. First, we scale the abundances of Si, Fe, and Mg by the same factor as the O abundance to account for metallicity. Then, we scale these elements by an additional factor of $(1-\xi)/(1-\xi_{\rm CLOUDY,ISM})$, where $\xi_{\rm CLOUDY,ISM}$ is the default CLOUDY ISM dust-to-metal ratio of 0.46.  We do not correct the C or O abundances for depletion; we specify these elements separately with the adopted metallicity and C/O ratio. 

We calculate nebular line EWs using the emergent line intensity and emergent stellar and nebular continua, including the effects of dust attenuation. These EWs therefore correspond to observed, rather than intrinsic values. The continuum value represents the average over four wavelength bins, two on each side of the emission line. Since absorption strongly affects the continuum level in the vicinity of the \lya\ line, for \lya\ we extend our estimate of the continuum level to include six wavelength bins or 40 \AA\ on each side of the line. In contrast with the EWs, we give all line ratios as intrinsic values; these ratios would correspond to observed values after applying corrections for dust extinction.

\subsection{Shock Models}
\label{sec:models:shocks}
Even in star-forming galaxies, shock emission can generate a substantial fraction of the nebular emission \citep[\eg][]{rich14, krabbe14}. Therefore, we explore adding shock emission to our pure photoionization models. The line ratios in the shocked gas come from the Mappings III shock models \citep{allen08} at the metallicity of the Small Magellanic Cloud ($Z\sim0.003$). We consider the total emission from shock and precursor gas for models with shock velocities of 125 to 1000 \kmps\ and magnetic field strengths of 0.5-10 $\mu$G. All models at $Z\sim0.003$ in the Mappings III shock model library have a density of 1 cm$^{-3}$. 

For each shock model, we scale the shock emission and add it to the predicted emission from the photoionization models such that shocks contribute 10\%, 50\%, or 90\%\ of the total H$\alpha$ emission:
\begin{equation}
f_{\rm shocks}=\frac{{\rm H}\alpha_{\rm shocks} \times\ s}{{\rm H}\alpha_{\rm shocks} \times\ s + {\rm H}\alpha_{\rm photo}},
\end{equation}
where $f_{\rm shocks}$ is the fractional shock contribution, $s$ is the scale factor applied to the shock emission, and the subscripts ``photo" and ``shock" indicate the emission from photoionization and shocks, respectively. We then derive the total emission from a given spectral line, $I_{\rm tot}$, as
\begin{equation}
I_{\rm tot} = I_{\rm shocks} \times\ s + I_{\rm photo}.
\end{equation}
We address the results from the shock$+$photoionization models in \S~\ref{sec:results:shocks}.

\subsection{Fiducial Model Parameters}
\label{sec:models:fiducial}
In the discussion below, unless otherwise specified, we assume fixed values for the density, C/O abundance, dust-to-metals ratio, geometry, optical depth, and shock contribution. We summarize these values in Table~\ref{table:fiducial}. For the hydrogen density, we adopt a default value of log~$n_{\rm H}$=2, which matches the estimated densities of low-metallicity emission-line galaxies \citep{jaskot13}. We set our fiducial C/O ratio to 0.2, consistent with observations of low-metallicity galaxies \citep{stark14}, and adopt the lower dust-to-metals ratio of 0.1. Finally, our fiducial model consists of the optically thick, filled sphere geometry and assumes no emission from shocked gas.

\begin{table*}
\vspace*{-0.2in}
\begin{center}
\caption{Fiducial Model Parameters}
\label{table:fiducial}
\begin{tabular}{lc}
\hline
Parameter & Value  \\ 
\hline
Input SED Model & BPASS, Instantaneous Burst \\
$n_{\rm H}$ & 100 cm$^{-3}$ \\
C/O & 0.20 \\
$\xi$ & 0.1 \\
Geometry & sphere, $f_S$=0.03 \\
$f_{{\rm H}\beta}$ & 1.0 \\
$f_{\rm shocks}$ & 0.0 \\
\hline
\end{tabular}
\end{center}
\end{table*}

\section{Physical Origin of Strong \ciii\ Emission}
\label{sec:results:c3}
\subsection{Effect of Ionizing SED}
\label{sec:results:sed}
The \ciii\ line originates almost entirely from collisional excitations. For the models with $Z<0.008$, collisional excitations account for an average of $100\%$ of the total \ciii\ emission. In \hii\ regions, stellar ionizing photons dominate the gas heating, and the stellar SED will therefore regulate \ciii\ excitation as well as ionization of C$^{+}$. 

\begin{figure*}
\plotone{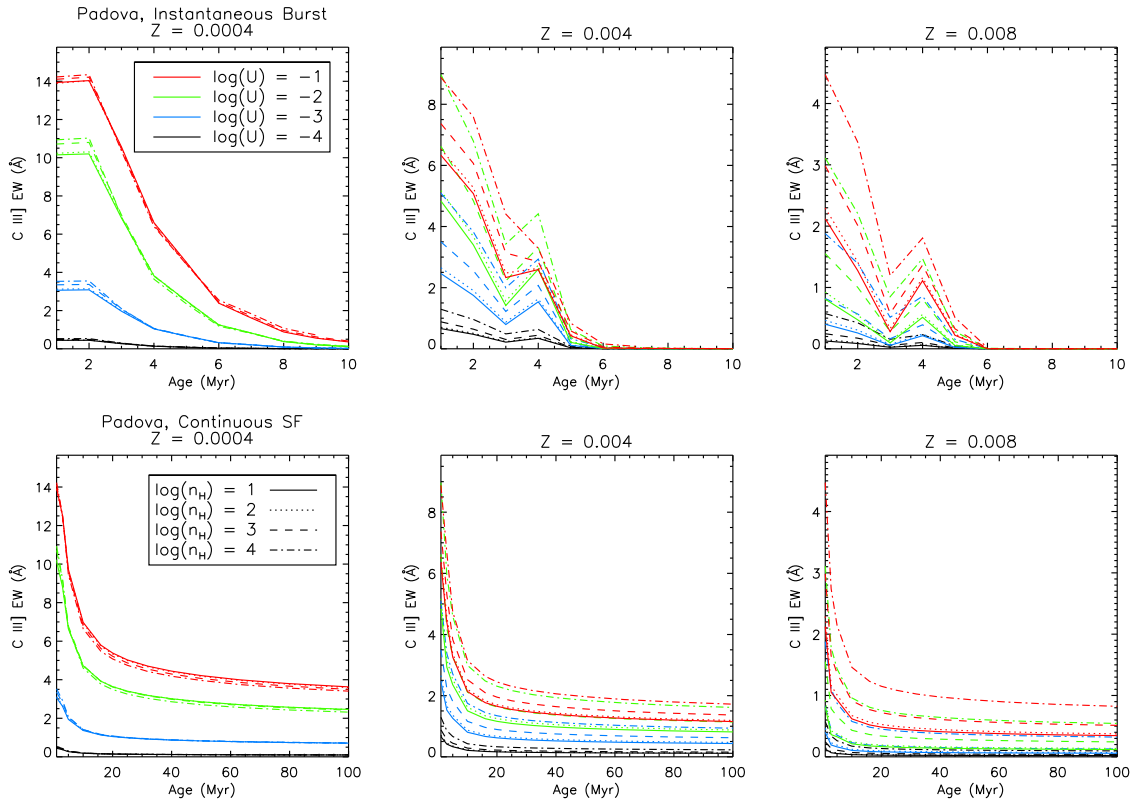}
\caption{The predicted \ciii\ EWs from the Padova input SEDs for different ages, metallicities, ionization parameters, and densities. Other parameters are as specified in Table~\ref{table:fiducial}. The upper panels show predictions for an instantaneous burst of star formation after different ages; the lower panels show predictions for continuous star formation at various ages. Each panel corresponds to a different metallicity for the stars and gas. Color indicates ionization parameter, and the line style indicates density.}
\label{fig:ew_pi}
\end{figure*}

\begin{figure*}
\epsscale{1.2}
\plotone{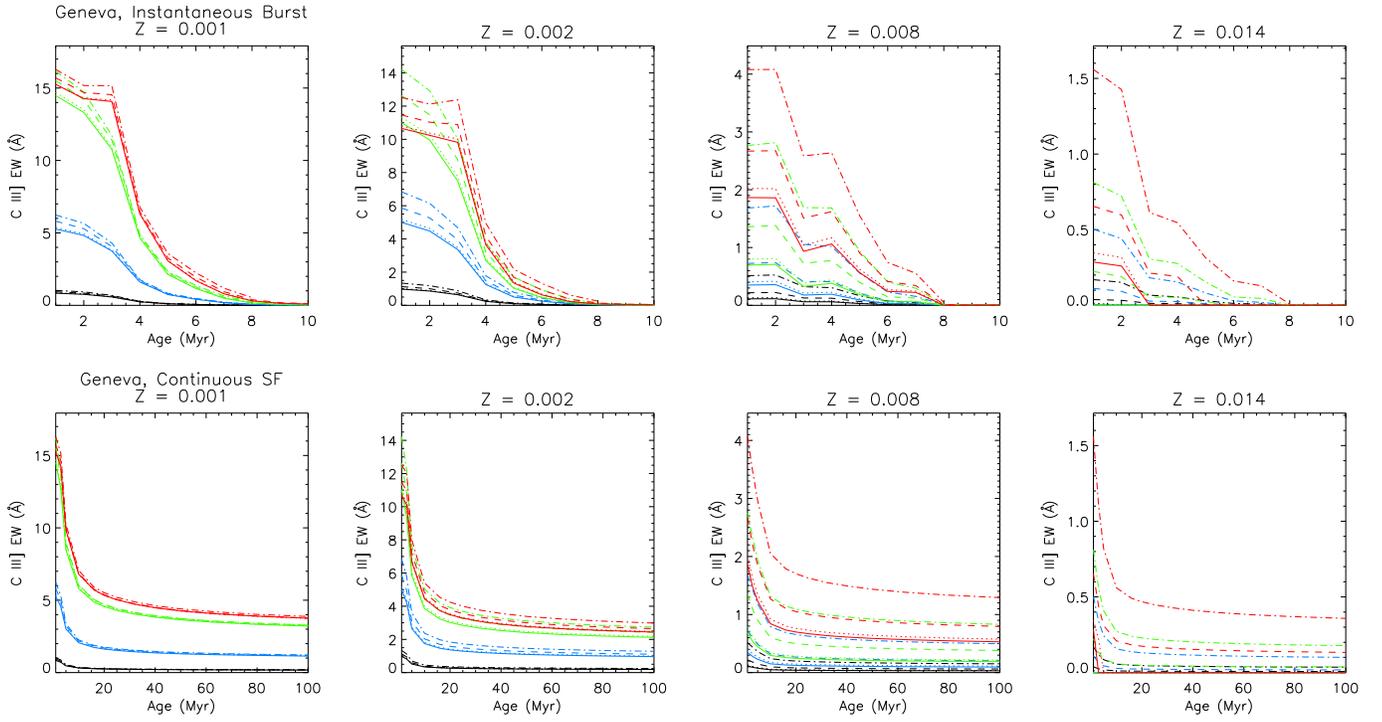}
\caption{Same as Figure~\ref{fig:ew_pi}, but for the predicted \ciii\ EWs from the Geneva input SEDs.}
\label{fig:ew_giv}
\end{figure*}

\begin{figure*}
\epsscale{1.2}
\plotone{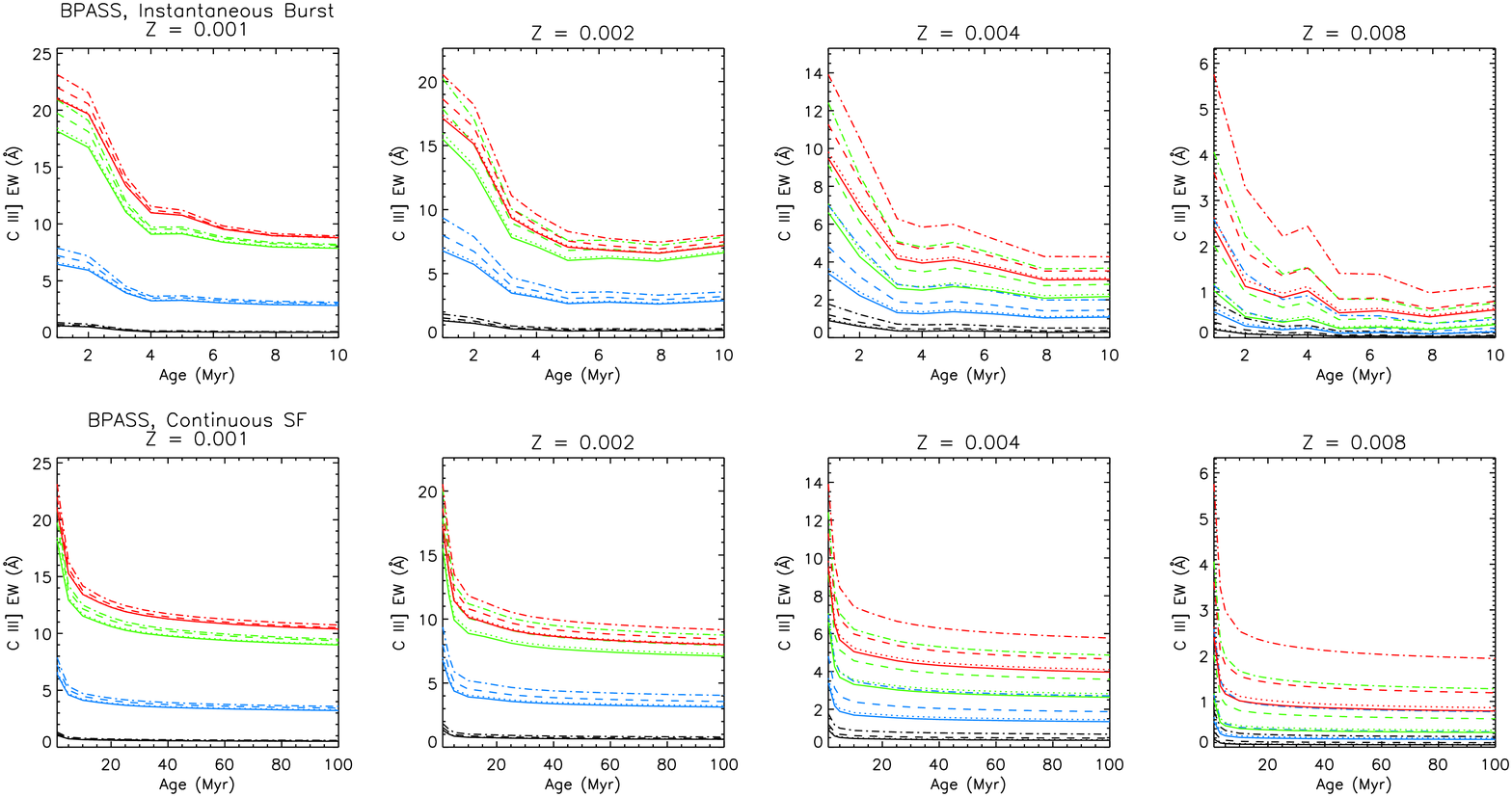}
\caption{Same as Figure~\ref{fig:ew_pi}, but for the predicted \ciii\ EWs from selected BPASS input SEDs.}
\label{fig:ew_bpass}
\end{figure*}

Figures~\ref{fig:ew_pi}-\ref{fig:ew_bpass} compare the \ciii\ EWs predicted at different ages for the Padova, Geneva, and BPASS stellar population models, respectively, with both instantaneous burst and continuous star formation histories. For each of the stellar population models, the expected \ciii\ EWs peak at the youngest ages ($<3$ Myr). The \ciii\ EW remains strong throughout the Wolf-Rayet phase, which peaks near 4 Myr after a burst for the $Z=0.004-0.008$ Padova models (Figure~\ref{fig:ew_pi}) and the  $Z=0.001-0.002$ Geneva models (Figure ~\ref{fig:ew_giv}). In reality, at ages $\lesssim3$ Myr, the surrounding neutral gas and dust may strongly obscure optical and UV emission from the starburst \citep[\eg][]{kobulnicky99, gilbert07, whitmore14, sokal15}, before feedback clears away some of this natal material. Therefore, the peak \ciii\ emission may occur instead at the slightly older ages corresponding to the Wolf-Rayet phase, consistent with the \ciii\ detections at low redshift \citep{rigby15}.

The effects of binary evolution result in both higher EW \ciii\ emission and in strong \ciii\ emission over a longer timescale. Only the BPASS models (Figure~\ref{fig:ew_bpass}) reach \ciii\ EWs $>17$\AA, high enough to explain the highest \ciii\ detections at low metallicity \citep{rigby15}. At the youngest ages, the harder spectrum of the BPASS models produces slightly higher gas temperatures and C$^+$ ionization rates relative to the other models and also predicts a weaker non-ionizing UV continuum at a given $U$. Binary interactions also provide an alternative channel for Wolf-Rayet star formation and can form Wolf-Rayet stars at later ages than predicted for single star evolution \citep{eldridge08, stanway16}. The extra Wolf-Rayet stars, secondary stars that have accreted mass, and binary merger products lead to an enhanced ionizing flux \citep{stanway16}. Consequently, the BPASS models maintain elevated \ciii\ EWs for a longer period of time relative to the single-star models. 

Models with a continuous star formation history (Figures~\ref{fig:ew_pi}-\ref{fig:ew_bpass}) paint a similar picture. \ciii\ emission is strongest at the youngest ages and for the BPASS models. At a fixed ionization parameter (i.e., fixed ionizing flux), the \ciii\ luminosity remains approximately constant with age. However, as the stellar SED becomes less hard at later ages, the same ionizing flux corresponds to a stronger non-ionizing UV continuum. As a result, the 1909\AA\ continuum level is higher and the \ciii\ EW is weaker at older ages. In contrast with the instantaneous burst models, the \ciii\ EWs in the continuous star formation models change little after 20 Myr. At this time, the population reaches an equilibrium between the birth and death of the most massive stars; only the increasing 1909 \AA\ continuum flux from the growing stellar population lowers the \ciii\ EW.

Such prolonged star formation could enrich the surrounding gas and raise the metallicity of any new stars. However, any change in metallicity also depends on the relative rates and metallicities of gas inflows and outflows. For instance, enriched outflows and metal-poor inflows may maintain relatively low metallicities in low-mass galaxies \citep[\eg][]{dalcanton07, dayal13}.

Binary stellar evolution effects appear crucial to understanding the high prevalence of strong \ciii\ emission among high redshift galaxies. For the rest of our analysis, we therefore focus on the predictions from the instantaneous burst BPASS models. 

\subsection{Effects of Nebular Parameters}
\label{sec:results:neb}
Collisional excitation rates and the abundance of C$^{2+}$ ions in the nebula ultimately determine the \ciii\ emission strength. Figure~\ref{fig:ew_bpass} illustrates the predicted \ciii\ EWs for the instantaneous burst BPASS ionizing SEDs with various ionization parameters, gas densities, and metallicities. 

The \ciii\ EW increases with $\bar{U}$ at all metallicities, as more photons can ionize C$^+$ and can heat the nebula through photoionization. In addition, the higher $\bar{U}$ nebulae have a larger physical size, which results in a higher gas column and higher extinction. Dust attenuates a larger fraction of the 1909 \AA\ continuum emission relative to the lower $\bar{U}$ models, providing an additional boost to the emergent \ciii\ EW. 

\ciii\ EWs also increase weakly with density. For fixed $\bar{U}$, a higher density nebula will have a higher incident ionizing flux (see Equation~\ref{eqn:u}). Consequently, the higher density models have slightly higher temperatures as well as higher collision rates and show enhanced \ciii\ emission. However, the magnitude of the effect is small; increasing the density by three orders of magnitude typically increases the EW by only a few {\AA}ngstroms.

Metallicity affects \ciii\ emission in three ways: by determining the shape of the ionizing SED, by influencing the nebular gas temperature, and by setting the overall carbon abundance. The first two effects act to enhance \ciii\ emission at low metallicities, while the third effect will suppress \ciii\ emission at low metallicities. The harder SEDs at low metallicities will boost the supply of C$^+$-ionizing photons. Secondly, since forbidden emission from metal lines constitutes the dominant cooling mechanism in \hii\ regions, the reduced gas metallicity will produce a higher temperature nebula and raise the \ciii~$\lambda$1909 collisional excitation rate. 

Figure~\ref{fig:ew_bpass} demonstrates that the $Z=0.002$ and $Z=0.001$ models do indeed show the strongest \ciii\ emission, even though they have fewer C atoms present. In contrast with \citet{erb10}, we find that the lowest metallicity models, at $Z=0.001$, generally show the strongest \ciii\ emission; the greater electron temperature obtained in these models compensates for the reduced C abundance. The models also show the same metallicity threshold for \ciii\ emission observed in low- and high-redshift galaxies \citep{rigby15}. By $Z=0.006$, the \ciii\ EW only reaches above 5 \AA\ at extreme ionization parameters and the youngest ages.

These models all assume a constant C/O ratio of $\sim$0.2. If instead we vary the C/O ratio, the \ciii\ EW scales almost linearly with C/O. Since carbon itself acts as a key coolant in the nebula, the higher nebular temperatures at low C/O ratios partially offset the decrease in C abundance. We discuss the effects of C/O ratio in more detail in the Appendix.

Dust-to-metals ratios and nebular geometry have minor effects on the \ciii\ EWs. We discuss the effects of these parameters in the Appendix. 

\subsection{Effect of Optical Depth}
\label{sec:results:tau}
Density-bounded nebulae have a lower column density of surrounding gas and lack the outermost nebular layers of radiation-bounded systems. These optically thin nebulae will therefore be deficient in lower ionization gas, which originates predominantly in the outer nebular regions \citep[\eg][]{pellegrini12, jaskot13}. 

\begin{figure*}
\plotone{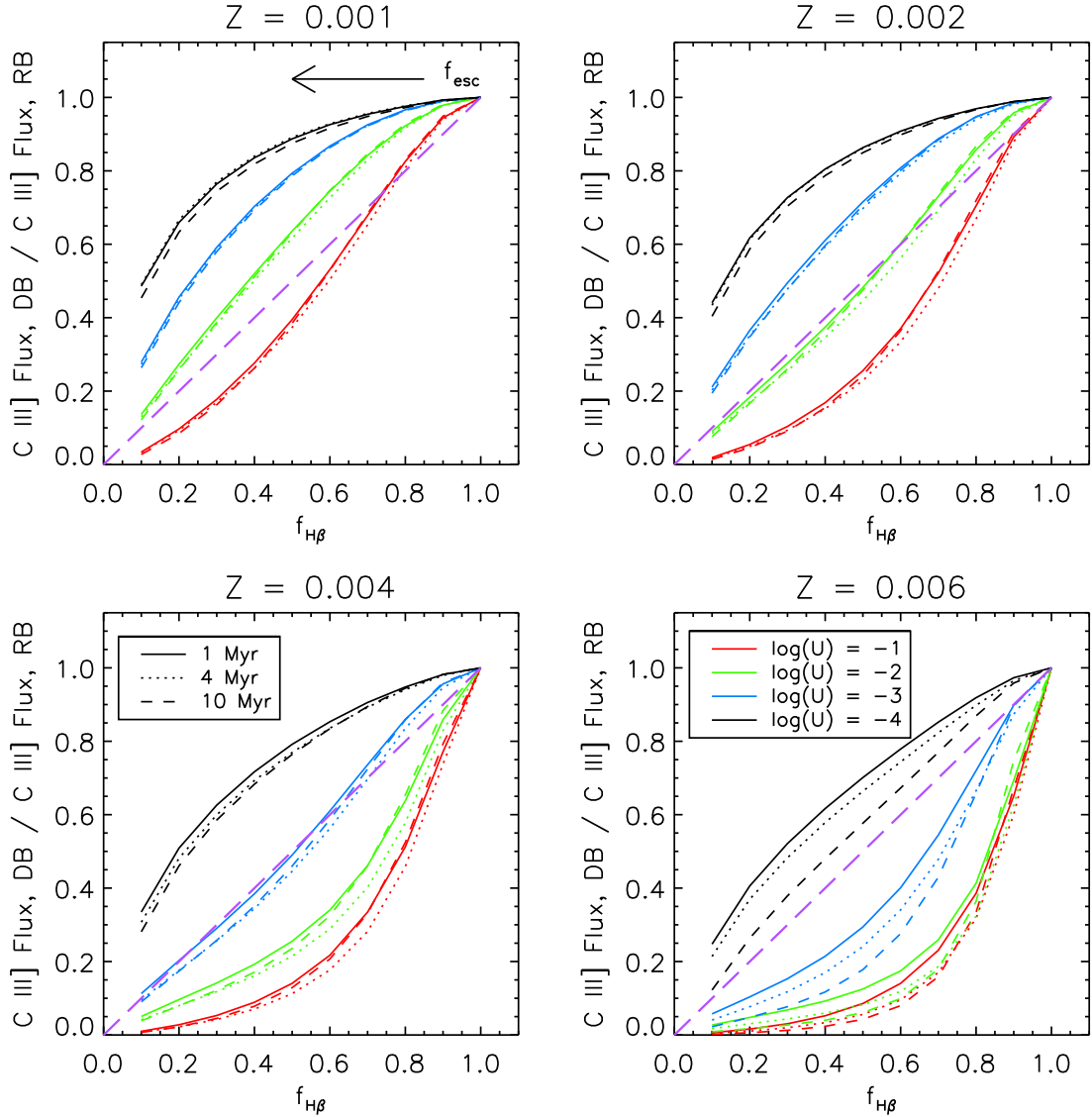}
\caption{Observed \ciii\ flux in a density-bounded scenario relative to the original \ciii\ flux in the optically thick, radiation-bounded case as a function of $f_{{\rm H}\beta}$ for the BPASS, instantaneous burst models. Each panel corresponds to a different metallicity. Line style denotes instantaneous burst age, and color indicates $\bar{U}$. Other parameters are as specified in Table~\ref{table:fiducial}. The long-dashed purple line shows a one-to-one relation, corresponding to a scenario in which the \ciii/H$\beta$ ratio remains constant with optical depth.}
\label{fig:c3_tau}
\end{figure*}

Figure~\ref{fig:c3_tau} shows that \ciii\ flux declines with decreasing $f_{{\rm H}\beta}$, i.e., truncating the nebular gas and decreasing the optical depth. The purple dashed line shows how the \ciii\ flux would change if both \ciii\ and H$\beta$ declined at the same rate with optical depth, in other words, if the \ciii/H$\beta$ ratio remained constant. The decline in \ciii\ flux with optical depth is steepest for the highest $\bar{U}$ models, where \ciii\ originates farther from the central source; in the inner nebular regions of the high $\bar{U}$ models, higher ionization \civ\ emission dominates over \ciii. For an escape fraction of ionizing radiation of $\sim$20\%, the value typically assumed for reionization by star-forming galaxies \citep[\eg][]{robertson15}, the \ciii\ flux may decrease by 2-68\% relative to the optically thick case. The lower ionization and lower metallicity models show less dramatic declines of \ciii\ emission with optical depth. 

\section{Relationship Between \ciii\ and UV and Optical Emission Lines}
\label{sec:results:linescalings}
\subsection{\lya}
\label{sec:results:lya}
\begin{figure*}
\plotone{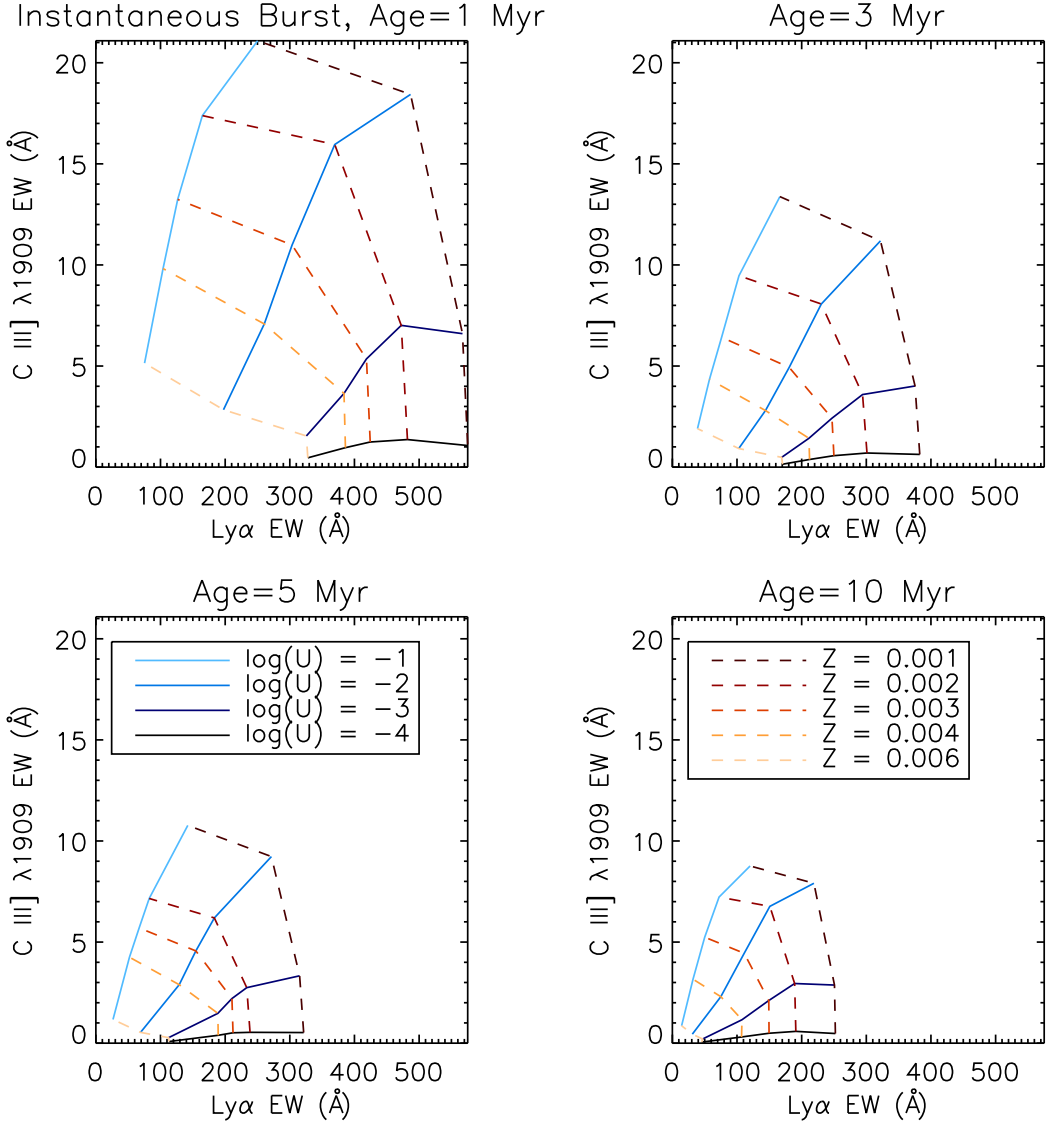}
\caption{\ciii\ EW vs. \lya\ EW. Each panel corresponds to a different instantaneous burst age. Solid lines indicate different $\bar{U}$ and dashed lines indicate different $Z$. Other parameters are as specified in Table~\ref{table:fiducial}.}
\label{fig:c3_lya}
\end{figure*}

Observationally, \lya\ correlates with \ciii\ in low- and high-redshift galaxies, although the scatter in this relation is high \citep{rigby15}. Our model grids show only a weak trend of \ciii\ with the nebular \lya\ produced in the starburst region; stellar \lya\ absorption should be negligible (4-8 \AA) for the young ages we consider here \citep[\eg][]{schaerer08, penaguerrero13}. In Figure~\ref{fig:c3_lya}, we show \ciii\ and \lya\ EWs for different metallicities, ionization parameters, and ages. At a given \lya\ EW, the scatter in \ciii\ EW can be as high as 10-20\AA, comparable to the observed level of scatter among galaxies \citep{rigby15}. The highest scatter appears for the lowest metallicity models; at higher metallicities, low nebular temperatures prevent high EW \ciii\ emission in almost all cases. 

In reality, radiative transfer effects from gas outside the \hii\ region can significantly alter the observed \lya\ EWs. The resonant \lya\ transition becomes optically thick at even low \hi\ column densities of $< N_{\rm HI}=10^{14}$ cm$^{-2}$ \citep[\eg][]{zheng02, verhamme15}, and scattering and dust extinction suppress \lya\ emission in many low-metallicity galaxies \citep[\eg][]{kunth94, atek09, thuan97, kunth98, james14}. The observed correlation between \ciii\ and \lya\ therefore suggests that conditions favorable for \ciii\ production are also favorable for \lya\ escape. 

\ciii\ emission should peak in galaxies during the Wolf-Rayet phase, when the star-forming regions generate intense ionizing flux and have cleared out surrounding absorbing material (see \S~\ref{sec:results:sed}). The strong ionizing flux at this time will also produce significant \lya\ emission, while early feedback from stellar winds and supernovae may clear out the pathways for \lya\ to escape. By shifting gas away from the galaxy's rest-frame velocity, these outflows may also lower the \lya\ optical depth at line center \citep[\eg][]{mashesse03}. 

Galaxies with high ionization parameters may likewise favor both \ciii\ formation and \lya\ escape. Highly concentrated star formation within a galaxy will result in a higher number of ionizing photons per surrounding atom and thus a high ionization parameter. The concentrated feedback from such systems may be more effective at clearing out material \citep[\eg][]{heckman11, alexandroff15} and may promote \lya\ escape, while the high $\bar{U}$ will also boost \ciii\ production (\S~\ref{sec:results:neb}). Indeed, galaxies with high ratios of \oiii/\oii, a measure of $\bar{U}$, often exhibit strong \lya\ emission \citep[\eg][]{nakajima14, jaskot14, henry15}. Finally, in lower metallicity environments, lower dust fractions could result in less \lya\ absorption, while higher electron temperatures and less dust will also allow greater production and transmission of \ciii.

\subsection{UV Emission Lines}
\label{sec:results:uv}
Consistent with the suggestion from \citet{stark14}, almost all of the photoionization models with $Z \leq 0.004$ predict that \ciii\ should be the next strongest emission line at $< 2700$ \AA\ after \lya. Emission from \oiiib~$\lambda\lambda$ 1661,1666 (hereafter \oiiib~$\lambda$1665) only becomes comparable to \ciii\ for models with both low C/O ratios (C/O $ \leq 0.12$) and high ionization parameters (log $\bar{U} \geq -2$). The different ionization potentials of the C and O lines lead to a dependence of \ciii/\oiiib\ on $\bar{U}$ as well as on the C/O abundance ratio \citep[c.f.,][]{erb10}. 

Similarly, the \heii~$\lambda$1640 line is weaker than \ciii. In fact, with a maximum nebular EW $< 1$ \AA, the models suggest that \heii~$\lambda$1640 will be difficult to detect in gas photoionized by star formation. Even at the lowest metallicity considered, $Z=0.001$, the \heii\ EW is less than 0.5 \AA. At low metallicities, the Geneva and Padova models likewise do not generate strong nebular \heii\ emission; in these models, \heii\ only becomes stronger than \ciii\ at moderate to high metallicities ($Z\geq0.004$) and during the Wolf-Rayet phase. Strong nebular \heii\ may therefore indicate gas ionized by shocks, non-stellar sources, or perhaps exotic Population III stars. The \heii\ detected in high-redshift galaxies often appears broad \citep[\eg][]{brinchmann08, erb10}, and likely originates from the winds of hot stars rather than the nebular gas.

In the near-UV, the photoionization models predict that certain low-ionization lines may exceed \ciii's emission strength in some cases. For instance, \hei~$\lambda$2945, \hei~$\lambda$3188, and \oii~$\lambda$2471 are stronger than \ciii\ for some models with log $\bar{U} \leq -3$, and \mgii~$\lambda\lambda$ 2796, 2803 emission appears stronger than \ciii\ in most models with log $\bar{U} <-1$. However, the photoionization models only probe the \hii\ region itself. In rest-UV spectra of star-forming galaxies, low-ionization resonant lines like \mgii\ often appear in absorption and trace gas outflows \citep[\eg][]{weiner09, rubin10}. The geometry and kinematics of the cool gas outside the \hii\ region may significantly alter both the observed \mgii\ strength and its line profile, and in low resolution spectra, absorption of \mgii\ in outflows could wash out the observed emission almost entirely \citep{prochaska11}.

The \ciii\ line is generally stronger than or comparable to the UV lines from other ionization states of carbon, such as the \civ~$\lambda\lambda$1548,1550 doublet (hereafter \civ~$\lambda$1549) or \ciip~$\lambda$2326. The higher ionization nebular \civ~$\lambda$1549 only dominates \ciii\ in some of the most extreme models with log~$\bar{U}=-1$ or with log~$\bar{U}=-2$ and C/O $=0.04$. Even in most of these cases, the ratio of \ciii/\civ\ is near unity (e.g., Figure~\ref{fig:c3ew_c4c2}). Conversely, \ciip~$\lambda$2326 becomes stronger than \ciii\ only in the lowest ionization parameter models with log~$\bar{U}=-4$ (Figure~\ref{fig:c3ew_c4c2}); such low values of $\bar{U}$ do not match the observed properties of most high-redshift galaxies \citep[\eg][]{brinchmann08, richard11, shirazi14, nakajima14}. Many of the low C/O ($\leq 0.12$) and high $\bar{U}$ (log $\bar{U} \geq -2$) models show that another \ciii\ line, \ciii~$\lambda$977 may emit more strongly than \ciii~$\lambda$1909. However, this line may partially overlap with \lyg\ absorption \citep{henry15}, which will make it difficult to observe. 

\subsubsection{Diagnostic Grids}
\begin{figure*}
\plotone{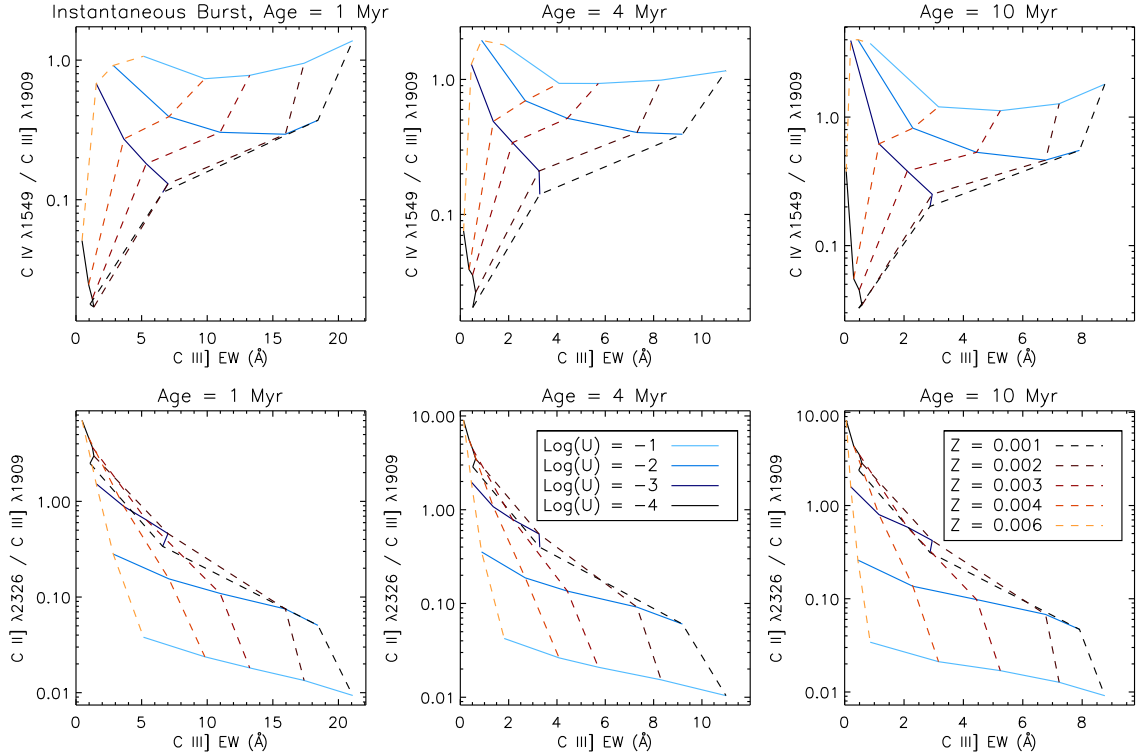}
\caption{Top row: Nebular \civ~$\lambda$1549/\ciii~$\lambda$1909 vs. \ciii\ EW. Bottom row: \ciip~$\lambda$2326/\ciii~$\lambda$1909 vs. \ciii\ EW. Each panel corresponds to a different instantaneous burst age. Solid lines indicate different $\bar{U}$ and dashed lines indicate different $Z$. Other parameters are as specified in Table~\ref{table:fiducial}.}
\label{fig:c3ew_c4c2}
\end{figure*}

If detected, the ratio of lines such as \ciip/\ciii\ and \civ/\ciii\ can constrain the ionization parameter; the \ciii\ EW will then set additional constraints on the starburst age and metallicity. Figure~\ref{fig:c3ew_c4c2} provides diagnostic grids of \ciii\ EWs and \cii/\ciii\ or \civ/\ciii\ as a function of age, $\bar{U}$, and $Z$. In practice, however, diagnostics using nebular \civ~$\lambda$1549 emission may prove difficult to employ. A strong stellar \civ\ P-Cygni profile will complicate the interpretation of \civ\ emission, and only high resolution spectra may successfully separate the stellar and nebular components \citep[\eg][]{crowther06b, quider09}. Depending on the kinematics and column density of interstellar gas, \civ\ absorption along the line-of-sight could also reduce observed \civ\ emission \citep{steidel16}. Nevertheless, recent detections of nebular \civ\ emission in lensed galaxies from $z=1.5-7$ suggest that nebular \civ\ may be observable in highly ionized galaxies and strong \ciii-emitters at high redshift \citep{stark14, stark15b}.
 
\subsection{Optical Emission Lines}
\label{sec:results:optical}

The same conditions responsible for strong \ciii\ emission will generate strong emission from other collisionally excited forbidden lines. High nebular temperatures, high ionization parameters, and strong ionizing radiation from young stellar populations will also enhance \oiii~$\lambda$5007 emission. Galaxies with extreme \oiii\ EWs appear common at $z>6$ \citep[\eg][]{labbe13, smit14}, and their rest-frame optical emission implies \oiii$+$H$\beta$ EWs as high as 1500 \AA \citep{smit14}.

\begin{figure*}
\plotone{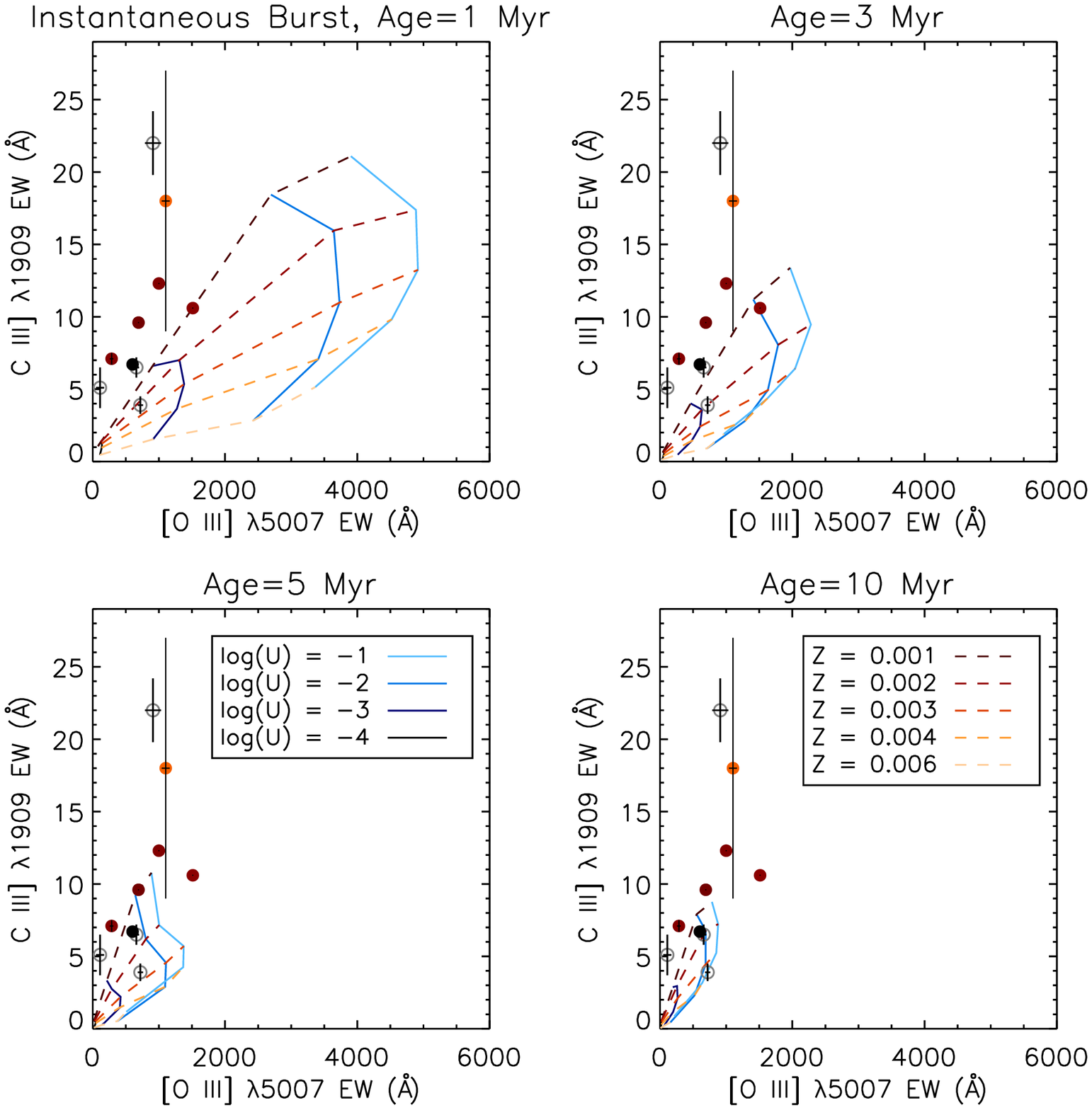}
\caption{\ciii\ EW vs. \oiii~$\lambda$5007 EW. Each panel corresponds to a different instantaneous burst age. Solid lines indicate different $\bar{U}$ and dashed lines indicate different $Z$. Other parameters are as specified in Table~\ref{table:fiducial}. The data points represent the observed EWs for low-mass, low-luminosity galaxies at $z\sim2$ from \citet{stark14}, the extreme \oiii~emitter Ion2 from \citet{debarros16}, and the low-metallicity galaxy at $z=2.3$ from \citet{erb10}. We include error bars when available. Open symbols indicate that the \oiii\ EW was derived from the best-fit model SEDs to broadband photometry \citep{stark14} and includes H$\beta$ emission. The open symbols therefore correspond to an upper limit on the \oiii\ EW. We color-code the galaxies based on their metallicity; the colors match the colors of the model grid metallicity lines. Galaxies plotted as gray circles do not have metallicity estimates. For clarity, we have omitted one outlier, G881-329, with highly uncertain parameters (\ciii\ EW$=7.1 \pm 3.1$ \AA; \oiii$+$H$\beta$ EW$=6230 \pm 3610$ \AA; \citealt{stark14}.}
\label{fig:c3_5007}
\end{figure*}

\begin{figure*}
\plotone{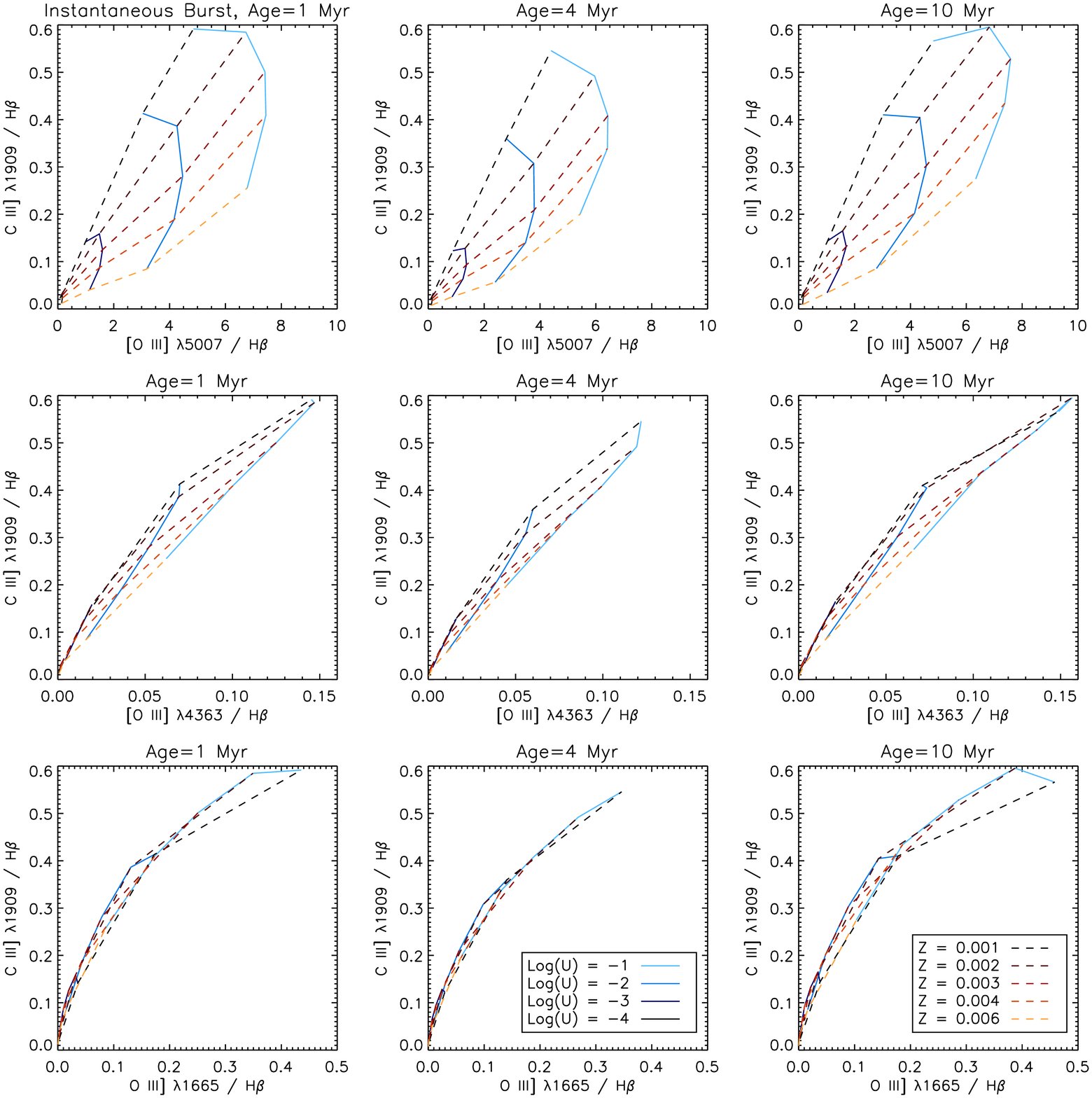}
\caption{Top row: \ciii/H$\beta$ vs. \oiii~$\lambda$5007/H$\beta$. Middle row: \ciii/H$\beta$ vs. \oiii~$\lambda$4363/H$\beta$. Bottom row: \ciii/H$\beta$ vs. \oiiib~$\lambda$1665/H$\beta$. Each panel corresponds to a different instantaneous burst age. Solid lines indicate different $\bar{U}$ and dashed lines indicate different $Z$. Other parameters are as specified in Table~\ref{table:fiducial}.}
\label{fig:c3_4363}
\end{figure*}

Figures~\ref{fig:c3_5007} and \ref{fig:c3_4363} show the predicted scalings between the \oiii~$\lambda$5007 and \ciii\ EWs and fluxes. The \ciii\ EW shows a much tighter correlation with \oiii\ EW than with the \lya\ emission from the \hii\ region, and both \oiii\ and \ciii\ EWs reach their highest values for young, high ionization parameter models. Through its dependence on temperature, the \ciii\ emission exhibits a greater sensitivity to metallicity than the \oiii\ emission. Figures~\ref{fig:c3_5007} and \ref{fig:c3_4363} only show models with a fixed C/O ratio; variations in the C/O fraction among galaxies will naturally lead to higher scatter between \ciii\ and \oiii\ emission. 

The \ciii\ flux shows an even tighter correlation with the temperature-sensitive \oiii~$\lambda$4363 line and \oiiib~$\lambda$1665 doublet (Figure~\ref{fig:c3_4363}). This correlation suggests that the typically weak \oiii~$\lambda$4363 line may reach a detectable level in galaxies with strong \ciii\ emission. We could then obtain metallicity estimates of these galaxies via the direct abundance method, a more reliable diagnostic than the alternative empirical strong-line methods. 

\subsubsection{Diagnostic Grids}
\begin{figure*}
\plotone{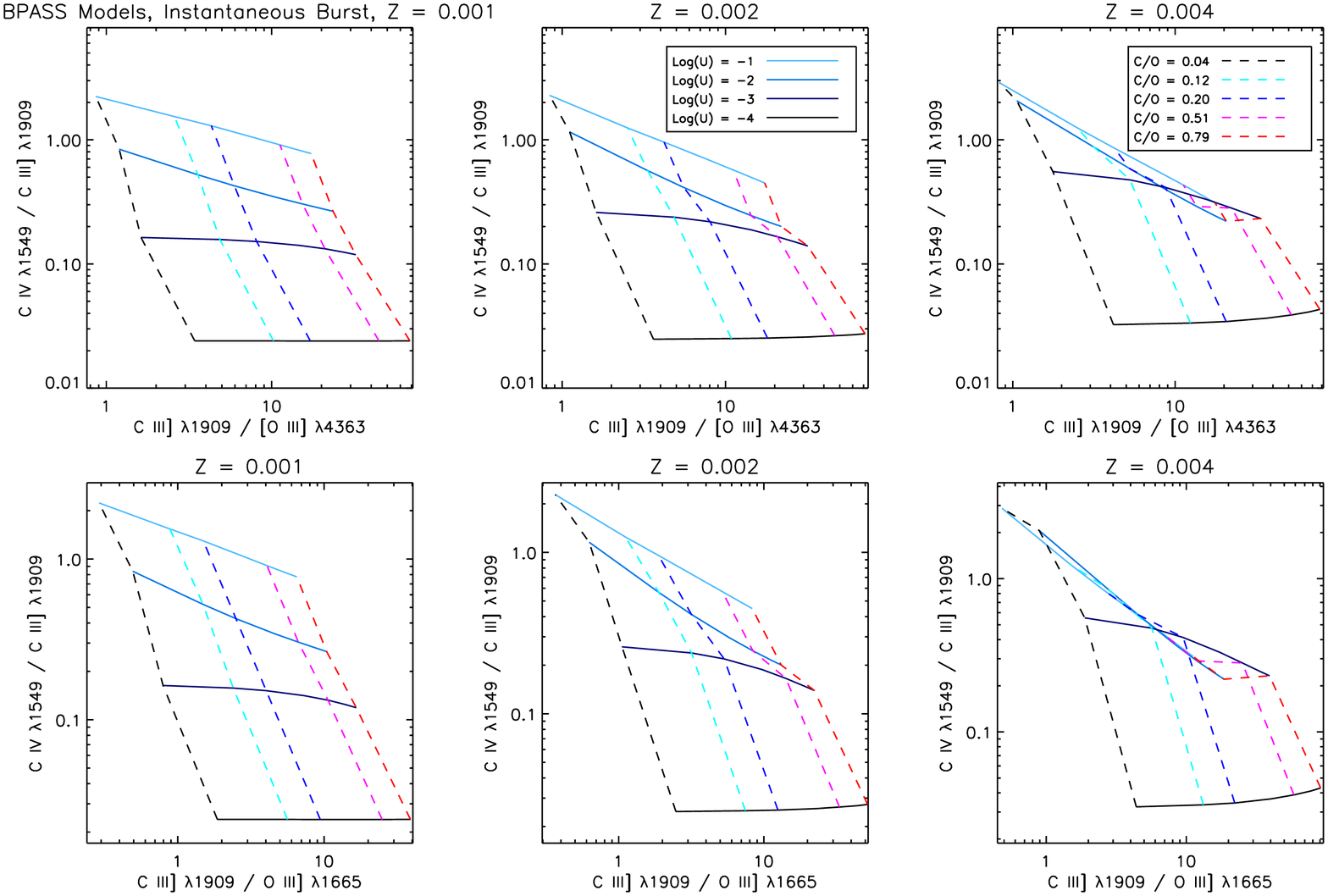}
\caption{Top row: \civ~$\lambda$1549/\ciii~$\lambda$1909 vs. \ciii~$\lambda$1909/\oiii~$\lambda$4363. Bottom row: \civ~$\lambda$1549/\ciii~$\lambda$1909 vs. \ciii~$\lambda$1909/\oiiib~$\lambda$1665. Each panel corresponds to a different metallicity. Solid lines indicate different $\bar{U}$ and dashed lines indicate different C/O ratios. The models shown correspond to an instantaneous burst age of 5 Myr. Ages of 1 or 10 Myr would shift the grids by $\leq$0.24 dex in \civ/\ciii, $\leq$0.06 dex in \ciii/\oiii~$\lambda$4363, and $\leq$0.09 dex in \ciii/\oiii~$\lambda$1665. Other parameters are as specified in Table~\ref{table:fiducial}.}
\label{fig:c4c3_co}
\end{figure*}

Together with \ciii, the \oiii~$\lambda$4363 and \oiiib~$\lambda$1665 lines can constrain nebular C/O ratios and dust extinction. The ratio of \ciii\ to \oiii~$\lambda$4363 flux shows little variation with age or metallicity and essentially depends almost entirely on the C/O ratio of the galaxy. As with \oiiib~$\lambda$1665, the ratio of  \ciii\ to \oiii~$\lambda$4363 does have a slight dependence on $\bar{U}$, as a result of the different ionization potentials of the C and O ions. Figure~\ref{fig:c4c3_co} illustrates diagnostic grids of $\bar{U}$ and C/O; the ratio of \ciii\ to \oiiib~$\lambda$1665 or \oiii~$\lambda$4363 traces C/O, while ratios such as \civ/\ciii\ or \cii/\ciii/ constrain $\bar{U}$. Although both \ciii/\oiiib~$\lambda$1665 and \ciii/\oiii~$\lambda$4363 set tight constraints on C/O, dust extinction will have a greater effect on the latter ratio. With estimates of C/O from \ciii/\oiiib~$\lambda$1665, the observed \ciii/\oiii~$\lambda$4363 ratio could therefore help constrain the relative extinction between optical and UV wavelengths.

\begin{figure*}
\plotone{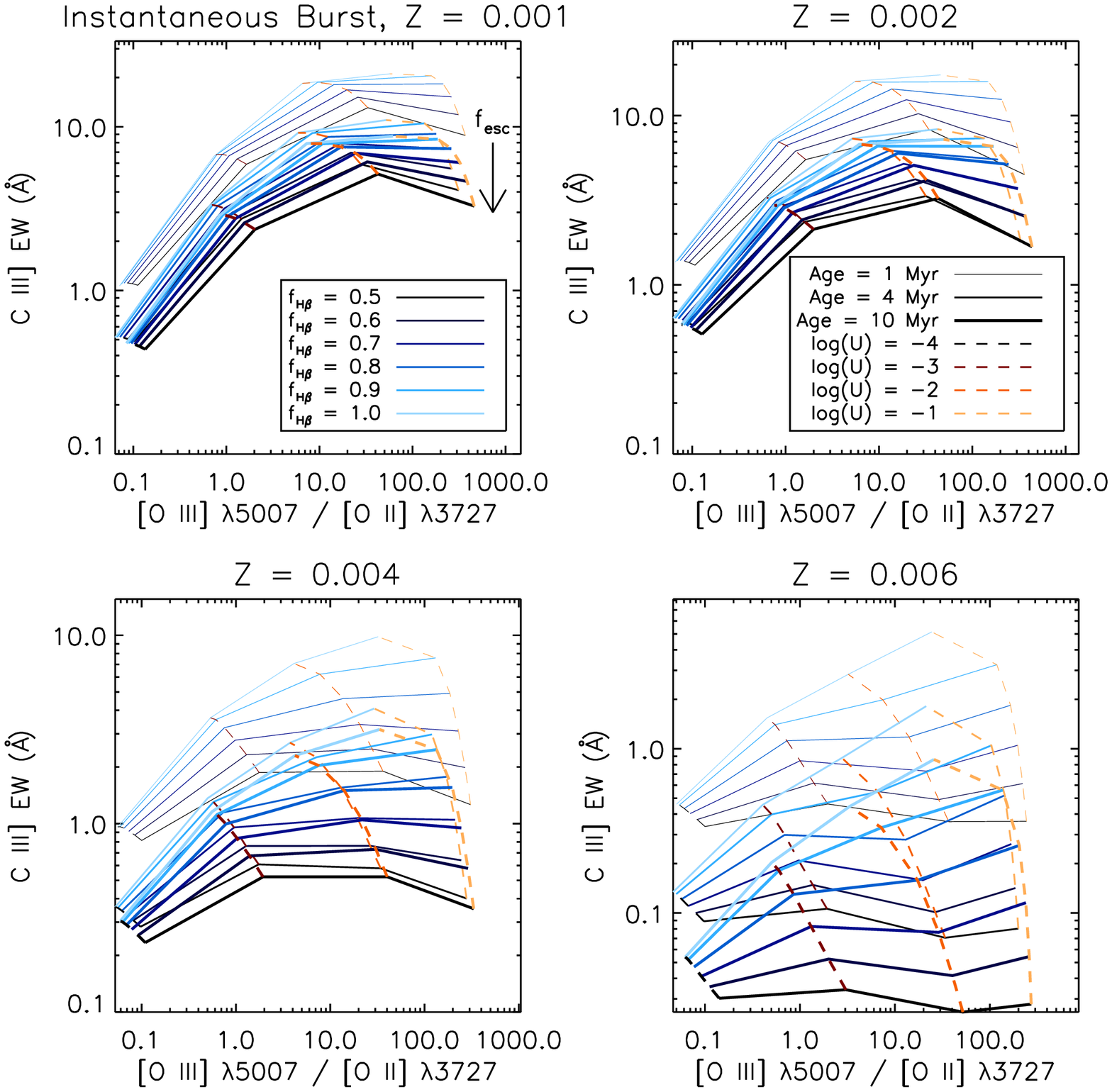}
\caption{\ciii~EW vs. \oiii~$\lambda$5007/\oii~$\lambda$3727. Each panel corresponds to a different metallicity. Solid lines indicate different optical depths $f_{{\rm H}\beta}$ and dashed lines indicate different $\bar{U}$. Thin, intermediate, and thick lines indicate instantaneous burst ages of 1, 4, and 10 Myr, respectively. Other parameters are as specified in Table~\ref{table:fiducial}.}
\label{fig:tau_diagnostic}
\end{figure*}

Since \ciii\ EWs decrease in density-bounded nebulae (\S~\ref{sec:results:tau}), \ciii\ could be a possible diagnostic of optical depth. \citet{bergvall13} predict similar behavior for optical emission lines, suggesting that weak nebular emission could select optically thin galaxies. Emission line ratios may also trace a galaxy's optical depth. \citet{jaskot13} and \citet{nakajima14} propose that galaxies with high \oiii/\oii\ ratios may have escaping LyC; density-bounded nebulae will lack the low-ionization gas at large radii that produces most of the \oii\ emission \citep[\eg][]{pellegrini12}. In Figure~\ref{fig:tau_diagnostic}, we plot grids of \oiii/\oii\ and \ciii\ EW as a function of age, $\bar{U}$, and optical depth. At a given metallicity, the \oiii/\oii\ ratio mostly depends on $\bar{U}$, with an additional weaker dependence on optical depth. However, optical depth strongly affects the \ciii\ EW, which can then break the degeneracy between $\bar{U}$ and optical depth. 

Figure~\ref{fig:tau_diagnostic} therefore suggests that, in combination with other emission lines, weak \ciii\ EWs could select density-bounded systems. The ratio of a high ionization and low ionization line, such as \oiii/\oii\ or \neiii/\oii, must first provide a constraint on $\bar{U}$. Galaxies with lower-than-predicted \ciii\ EWs at a given \oiii/\oii\ will then likely have high LyC escape fractions. Age also affects the \ciii\ EW by boosting the \ciii\ EW at ages $< $ 3 Myr; however, at ages of 4-10 Myr, the \ciii\ EW remains approximately constant. Likewise, metallicity also affects the scaling of \ciii\ EW with optical depth, and setting precise constraints on optical depth will require an estimate of metallicity.  Nevertheless, even without metallicity or age estimates, Figure~\ref{fig:tau_diagnostic} could select likely optically thin systems; for instance, a galaxy with \oiii/\oii$=10$ and a \ciii\ EW $\leq4$ \AA\ falls in the optically thin regime in all diagrams with $Z<0.004$. 

The behavior of \ciii\ EWs with optical depth is not unique; other emission lines should also show decreased EWs and could replace \ciii\ on the y-axis of Figure~\ref{fig:tau_diagnostic}. However, continuum emission from older stellar populations will also decrease the EWs of optical nebular emission lines and will complicate their interpretation. Higher ionization lines, such as \heii~$\lambda$1640, which are produced closer to the ionizing source, will not decrease with optical depth. In fact, \heii\ EWs could increase at low optical depth due to a weaker nebular continuum \citep{raiter10}. The EWs of moderate ionization UV lines such as \ciii\ are therefore preferable as optical depth diagnostics.

\subsection{Shocks and Shock Diagnostics}
\label{sec:results:shocks}

\begin{figure*}
\plotone{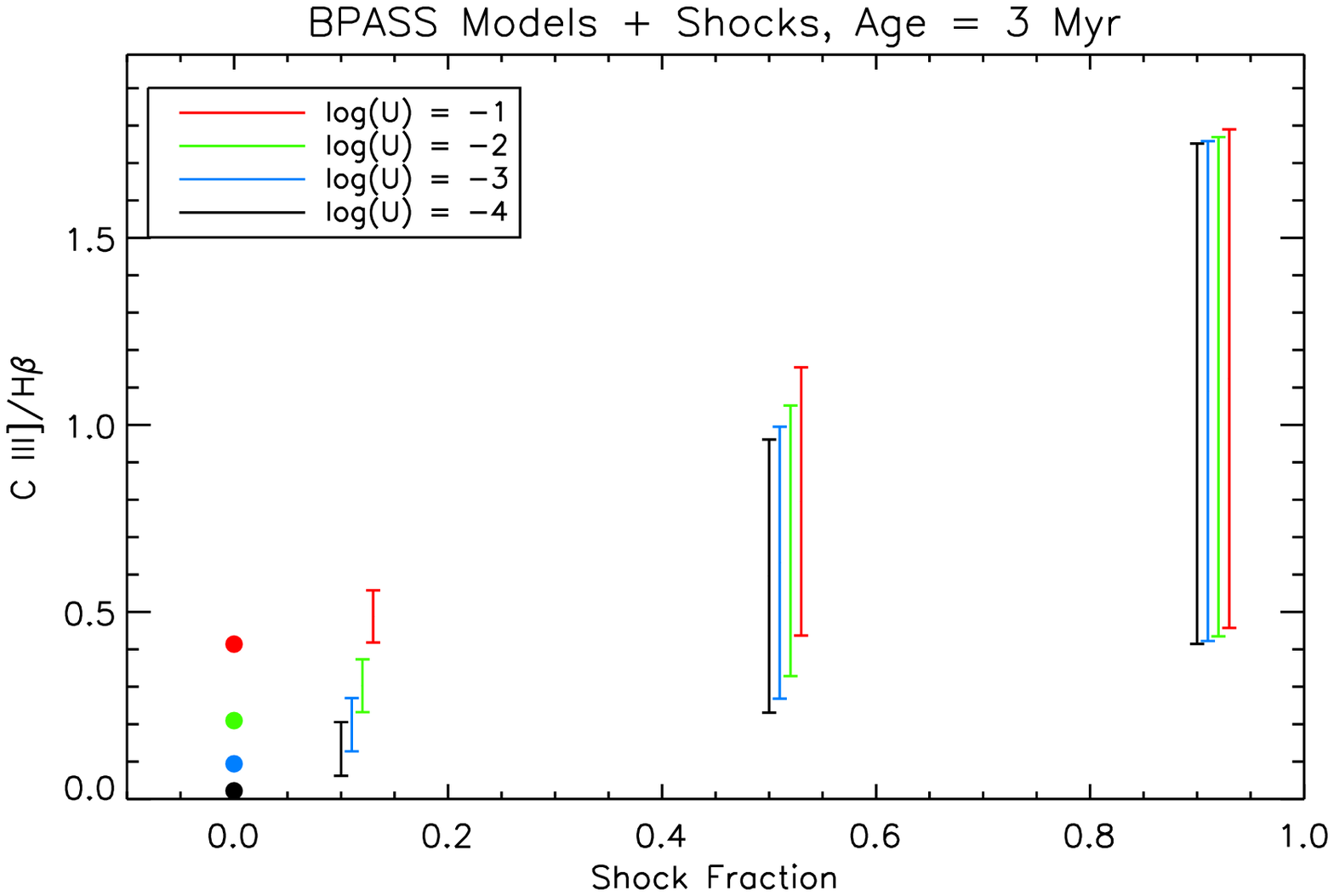}
\caption{\ciii/H$\beta$ vs. shock fraction. The filled circles show the \ciii/H$\beta$ ratio for pure photoionization with no shock emission. Vertical lines indicate the minimum and maximum \ciii/H$\beta$ for all shock models, with shocks contributing 10\%, 50\%, and 90\%\ of the total Balmer line emission. The lines are offset for clarity.  Color indicates the ionization parameter adopted for the photoionized gas. Other parameters are as specified in Table~\ref{table:fiducial}.}
\label{fig:shockrange}
\end{figure*}

If present, shock emission will almost always increase the observed \ciii\ flux. Figure~\ref{fig:shockrange} shows the change in \ciii/H$\beta$ when we add a shock emission contribution as described in \S~\ref{sec:models:shocks}. In nearly all cases, models with shock emission show stronger \ciii. Shock models with lower velocities and higher magnetic field strengths tend to produce the strongest \ciii\ emission. Shocks may have potentially significant effects on observed \ciii\ emission strengths. 

Although the \heii~$\lambda$1640 line appears weak in all the pure photoionization models, the shock models do predict appreciable \heii\ emission. The \heii\ flux steadily increases as we raise the fractional contribution of shocks to the observed emission. When shocks generate $\leq10$\% of the total Balmer line emission, the \heii~$\lambda$1640/H$\beta$ ratio is $\lesssim0.2$. For a shock contribution of 90\%, the \heii/H$\beta$ ratio rises to $>1$ in the majority of models. 

\begin{figure*}
\plotone{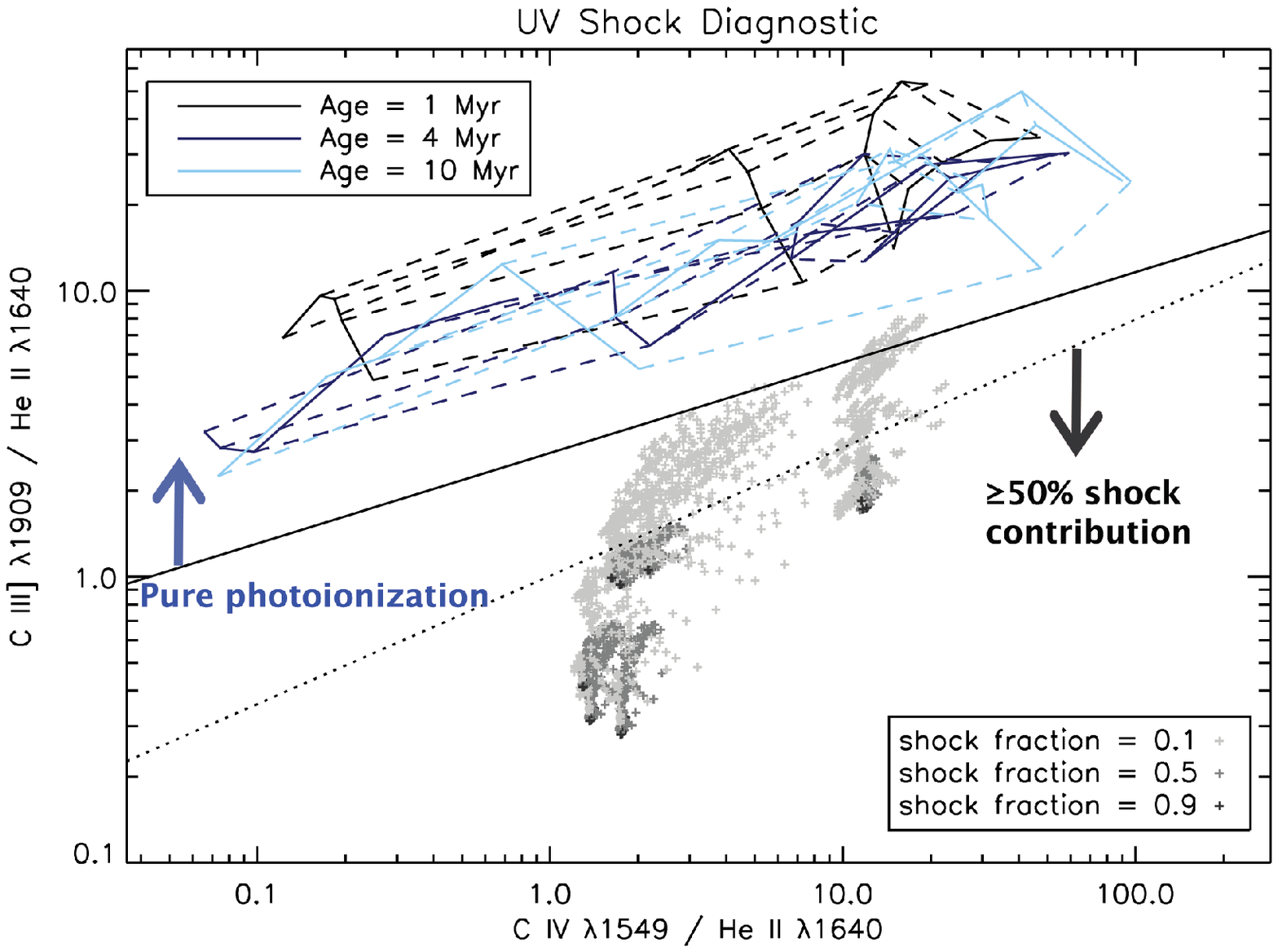}
\caption{The \ciii~$\lambda$1909/\heii~$\lambda$1640 vs. \civ~$\lambda$1549/\heii~$\lambda$1640 shock diagnostic. Solid and dashed lines show pure photoionization model grids; solid lines correspond to constant $\bar{U}$ and dashed lines correspond to constant metallicity ($Z=0.001-0.006$). We color-code each grid by burst age. Crosses show the predicted line ratios from the photoionization$+$shock models. Darker colors indicating a higher shock contribution. The shocked gas is at $Z=0.003$ in all models. For simplicity, we only plot predictions for the minimum, middle, and maximum shock velocity and shock magnetic field strength. The star-forming models all lie above the solid black line, and models where shocks dominate ($\geq 50$\% contribution) fall below the dotted line.}
\label{fig:shockdiagnostic}
\end{figure*}

Nebular \heii\ emission may therefore serve as a useful diagnostic of shocks in galaxy UV spectra. We display one possible diagnostic, \ciii/\heii\ vs. \civ/\heii\ in Figure~\ref{fig:shockdiagnostic}. \citet{villarmartin97} and \citet{allen98} proposed this particular diagnostic to separate shocks from AGN photoionization, while \citet{feltre16} has explored its use in distinguishing between AGN and star-forming galaxies. Here, we show that this diagnostic separates gas photoionized by star formation from emission dominated by shocks. The star-forming models lie above the solid black line:
\begin{equation}
\label{eqn:shocks_sf}
y=0.43+0.32 x,
\end{equation}
where $y$ is log (\ciii~$\lambda$1909 / \heii~$\lambda$1640) and $x$ is log (\civ~$\lambda$1549 / \heii~$\lambda$1640). 
Most of the models with a shock contribution of only 10\%\ also fall below this line, which illustrates the sensitivity of \heii~$\lambda$1640 emission to even low fractions of shock emission. The dotted line,
\begin{equation}
\label{eqn:shocks_strong}
y=0.003+0.45x,
\end{equation}
marks the limit where shock emission dominates; the models with shock contributions of $\geq50$\%\ lie in the region of the plot below this line. Although the shock models span a wide range in velocity and magnetic field strengths (\S~\ref{sec:models:shocks}), we caution that they only cover a single metallicity ($Z=0.003$) and density ($n=1$ cm$^{-3}$). The diagnostics in Equations~\ref{eqn:shocks_sf} and \ref{eqn:shocks_strong} may not accurately separate photoionized and shocked gas for higher densities or metallicities other than $Z=0.003$.

\section{Comparison with Observations}
\begin{figure*}
\plotone{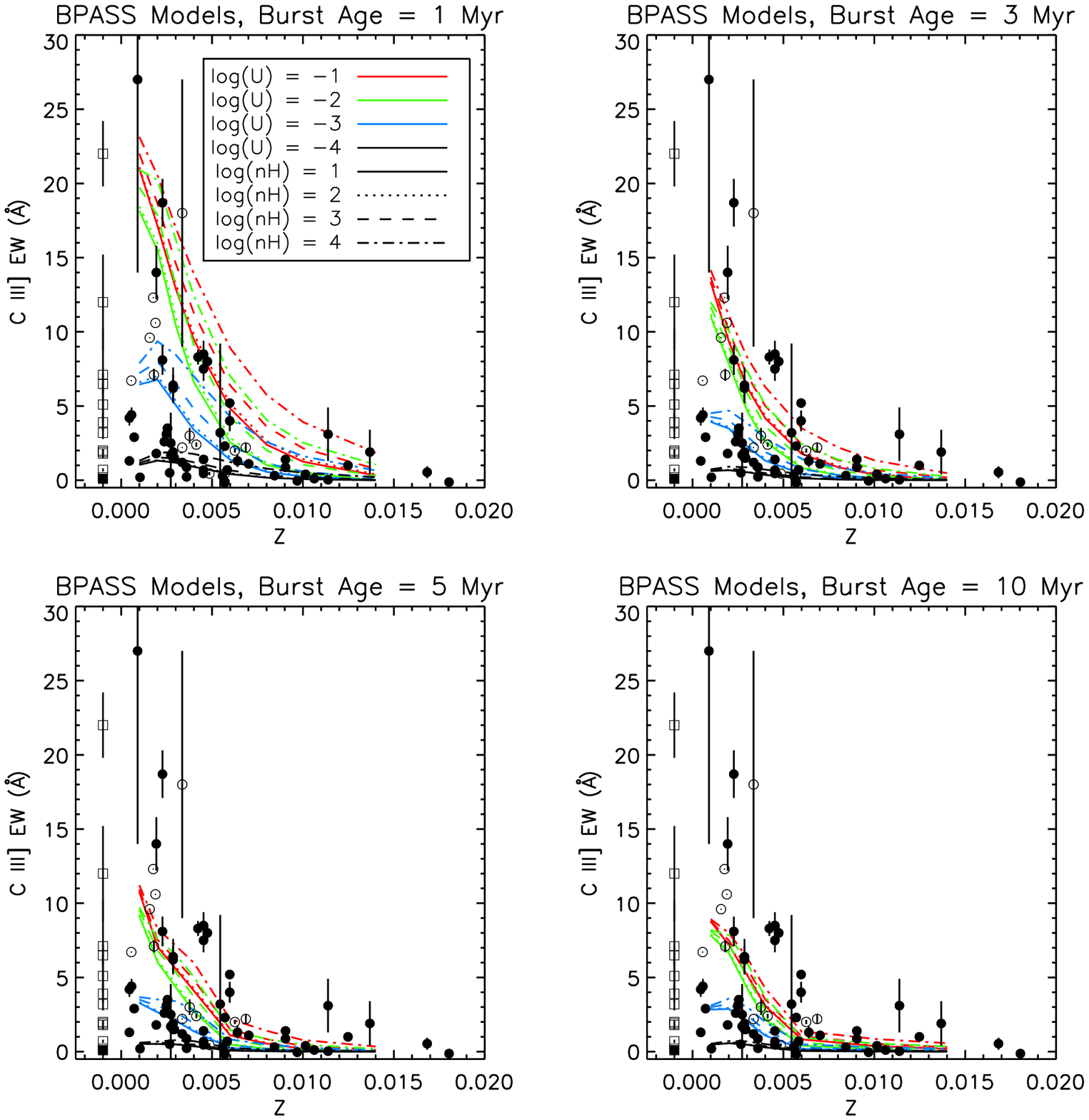}
\caption{Predicted \ciii\ EW as a function of $Z$. Each panel corresponds to a different instantaneous burst age. Model line styles and colors are the same as in Figure~\ref{fig:ew_pi}. Other parameters are as specified in Table~\ref{table:fiducial}. Filled data points show observed low-redshift galaxies, and open symbols show high-redshift galaxies \citep{rigby15, erb10, stark14, vanzella16b}. Squares indicate galaxies without published metallicity information; we arbitrarily plot these galaxies at an unphysical value of $Z=-0.001$. Error bars come from the published reference, where available.}
\label{fig:ew_z_obs}
\end{figure*}

Observed \ciii\ EWs from low- and high-redshift galaxies span a range of $\sim0$\AA\ to 27\AA\ \citep{rigby15}. In Figures~\ref{fig:ew_z_obs} and \ref{fig:c3_5007}, we compare the predictions of our photoionization models to observed \ciii\ EWs from the literature. The data shown include the compilation of \ciii\ EW measurements for the nearby ($z\sim0$) galaxies and distant ($z\sim1.6-3$) lensed galaxies from \citet{rigby15}. We also include the \ciii\ EW measurements for high redshift galaxies from \citet{erb10}, \citet{stark14}, and \citet{vanzella16b}. The observations we consider are a heterogeneous sample. The spectral data span a range of aperture sizes \citep{rigby15}, and some galaxies have multiple measurements, corresponding to the emission from different star-forming regions. Our models represent the emission from these individual star-forming regions; the emission from an integrated galaxy spectrum will be a luminosity-weighted average of the emission from multiple star-forming regions.

The BPASS models successfully reproduce almost the full range of observed \ciii\ EWs, as shown in Figure~\ref{fig:ew_z_obs}. Filled symbols indicate galaxies at $z<1$, and open symbols indicate galaxies at $z>1$. We arbitrarily plot galaxies without metallicity measurements at an unphysical negative metallicity of $Z=-0.001$. Low metallicity models are consistent with the high observed \ciii\ EWs and the galaxies' existing metallicity constraints.  At high metallicities, both the observed and predicted EWs decline steeply. 

In addition to low metallicity, most strong \ciii-emitting galaxies with EWs above 10 \AA\ require either a high ionization parameter (log $\bar{U}\geq-2$) or a young age ($\leq$3 Myr after an instantaneous burst or $\lesssim$10 Myr of continuous star formation; see also Figure~\ref{fig:ew_bpass}) to explain their observed \ciii\ EWs. For the youngest ages, even moderate ionization parameters can account for the observed \ciii\ emission. Values of log~$\bar{U}$ near $-2$ agree with the photoionization model estimates from \citet{stark14} for four high-redshift \ciii-emitting galaxies. However, our models suggest that the weakest modeled \ciii\ emitter in \citet{stark14}, with an EW of 6.7\AA, could even have a more moderate ionization parameter of log $\bar{U} < -2 $, given its low metallicity ($Z=0.0006$). 

Only one galaxy, Tol 1214-277, has \ciii\ emission higher than the fiducial models predict (EW$=27$\AA), but this galaxy's \ciii\ EW is poorly constrained. If Tol 1214-277's \ciii\ EW does indeed exceed 23 \AA, a higher C/O abundance (Figure~\ref{fig:ew_co}) or shock contribution (\S~\ref{sec:results:shocks}) could possibly account for its high EW. We have also only considered one choice of IMF. By increasing the ionizing flux at a given 2000 \AA\ continuum flux, a more top heavy IMF could raise the model \ciii\ EWs.

The models also imply that I Zw 18, with \ciii\ EWs of 1.3 to 4.4\AA\ in its different nebular regions, has a low ionization parameter of log $\bar{U}\sim -3$. This inferred ionization parameter conflicts with the known conditions in I Zw 18 (log~$\bar{U}=-2.5$ to $-2.2$; \citealt{dufour88, moralesluis14}), as already noted by \citet{rigby15}. A low C/O ratio ($<0.2$) could depress the \ciii\ EW in I Zw 18, but the galaxy's measured C/O ratio does not appear low, with a value of $\sim$0.17-0.25 \citep{garnett97, izotov99}. Density-bounded \hii\ regions within the galaxy could reduce the \ciii\ EW (\S~\ref{sec:results:tau} and \S~\ref{sec:results:optical}). Alternatively, dust extinction may play a role. Dust may lower nebular EWs by preferentially attenuating the ionizing continuum, thereby reducing emission line fluxes more than the underlying stellar continuum \citep[\eg][]{calzetti97, charlot00}. Dust extinction ranges from $A_V=0-0.47$ mag in I Zw 18 \citep{cannon02} and could affect the \ciii\ in some regions. Nevertheless, the cause of the weak \ciii\ in I Zw 18 remains unclear.

In Figure~\ref{fig:c3_5007}, we compare the \oiii\ EWs of the observed \ciii-emitting sample with the model predictions. For the weaker \ciii-emitters, ages of $\sim 5$ Myr) and moderate to high ionization parameters show the best agreement with the observations. However the grids do not simultaneously reproduce the \oiii\ and \ciii\ EWs of the strongest \ciii-emitters. The model \oiii\ EWs are the likely source of this discrepancy. The photoionization models do not include any continuum emission from an older stellar population, which will decrease the predicted EWs at optical wavelengths. This continuum emission will not contribute significantly at UV wavelengths and will not strongly affect the \ciii\ EWs.

\subsection{Predictions for Extreme Emission Line Galaxies}
Galaxies with low metallicities, high ionization parameters, and high optical emission line EWs have the greatest potential for strong \ciii\ emission. LAEs at $z=2-3$ fulfill these criteria; their high \oiii/\oii\ ratios imply a high ionization parameter \citep{nakajima14}, they have low metallicities $<0.3$\Zsol\ \citep{finkelstein11}, and they exhibit \oiii~$\lambda$5007 EWs that are typically $\geq\ 100$\AA\ \citep[\eg][]{hagen16}. The low-redshift GPs may represent local analogs of these high-redshift LAEs, with similarly strong \lya\ emission, high \oiii/\oii\ ratios, and low metallicities \citep{izotov11, jaskot13, jaskot14, nakajima14, henry15}. The GPs will provide a valuable test of photoionization models, as their optical spectra reveal numerous emission lines, many of which are difficult to observe at high redshift. 

\begin{figure*}
\epsscale{0.9}
\plotone{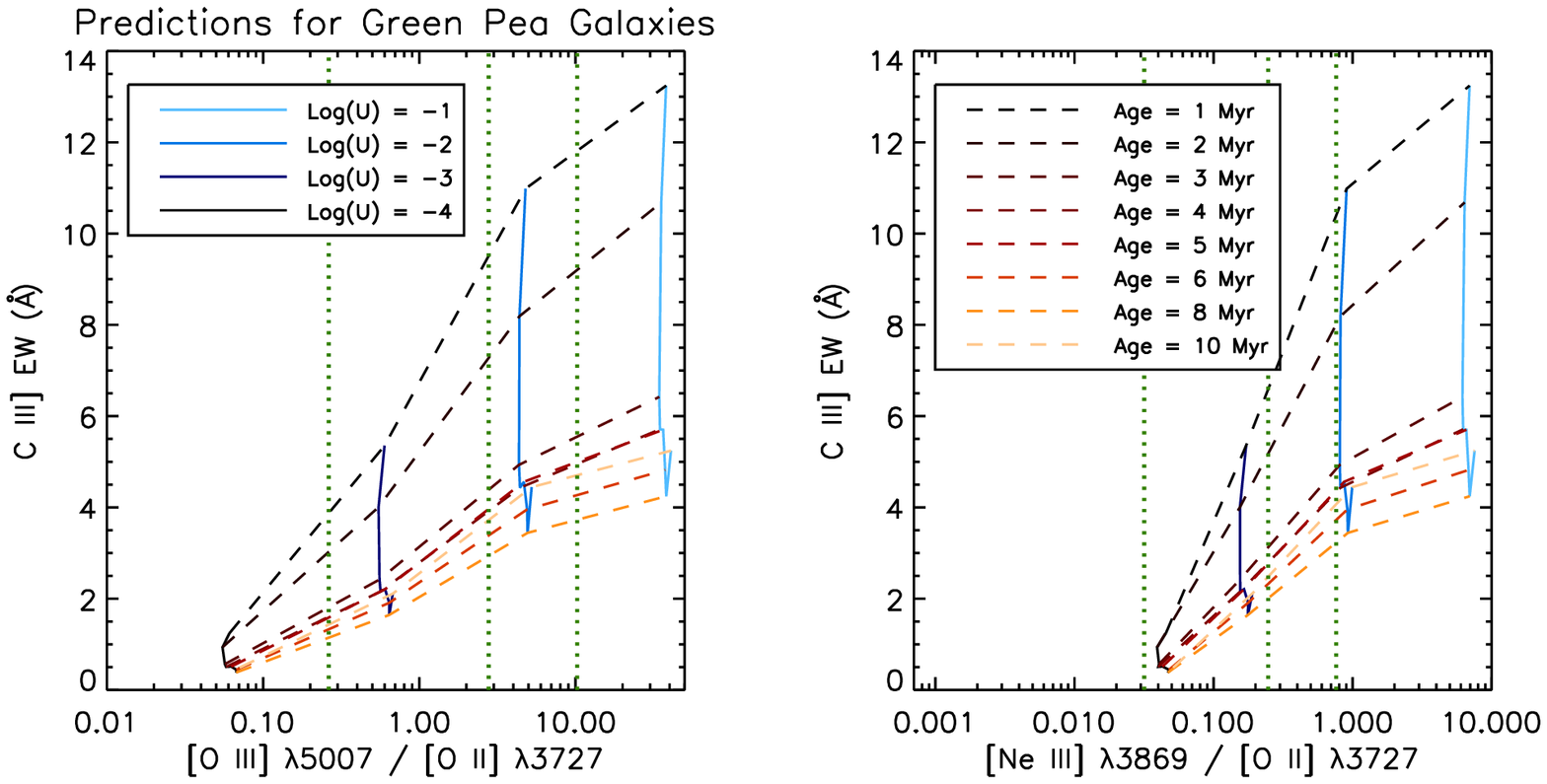}
\epsscale{0.7}
\plotone{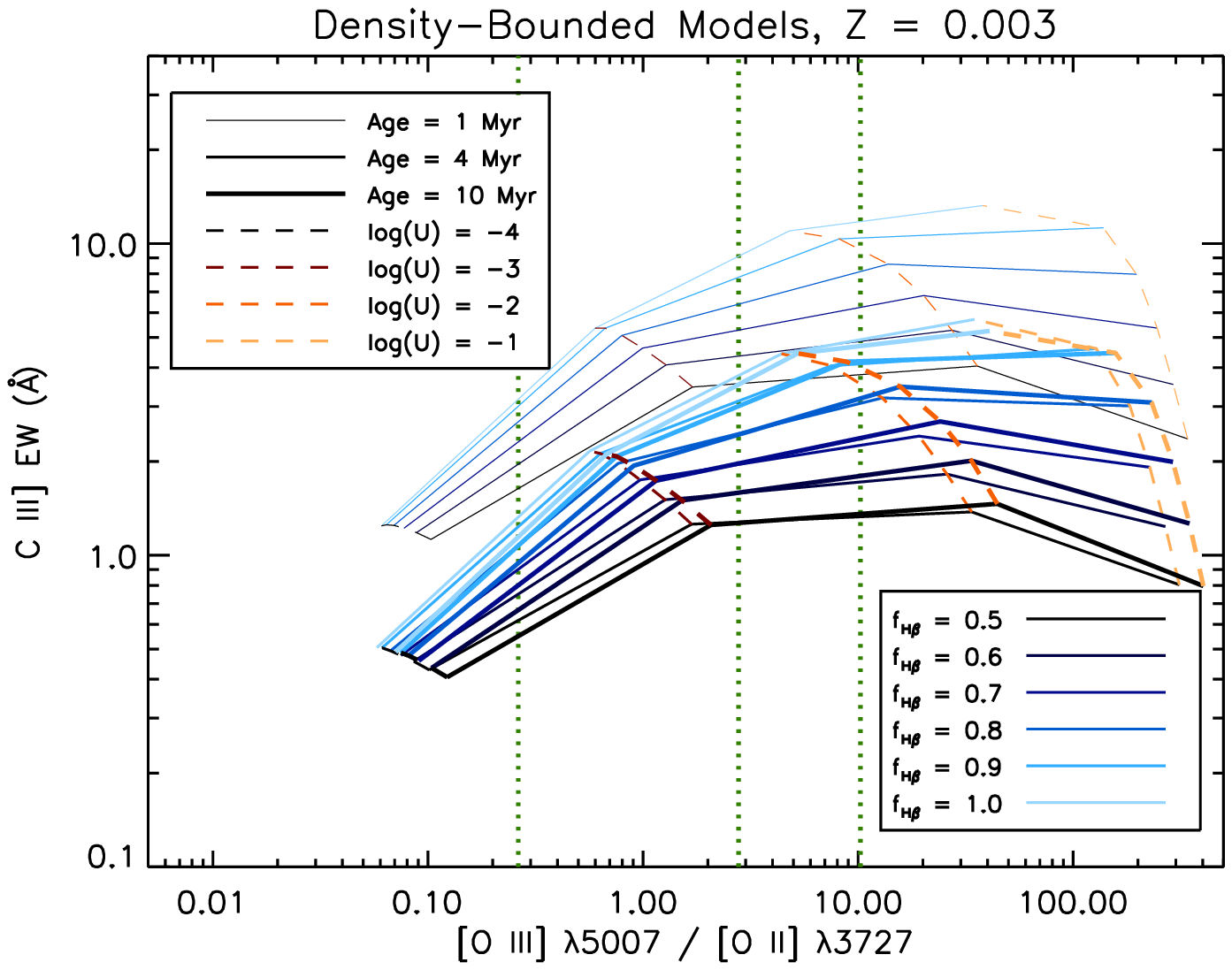}
\caption{Top left: \ciii\ EW vs. \oiii~$\lambda$5007/\oii~$\lambda$3727. Top right: \ciii\ EW vs. \neiii~$\lambda$3869/\oii~$\lambda$3727. Solid, blue lines indicate $\bar{U}$, and dashed lines indicate instantaneous burst age (1-10 Myr). Bottom: \ciii~EW vs. \oiii~$\lambda$5007/\oii~$\lambda$3727 as a function of $\bar{U}$ (dashed lines), optical depth $f_{{\rm H}\beta}$ (solid lines), and burst age (line thickness), as in Figure~\ref{fig:tau_diagnostic}. All models have a metallicity of $Z=0.003$, the average metallicity of the GP galaxies \citep[\eg][]{izotov11}. Other model parameters are as specified in Table~\ref{table:fiducial}. Vertical, dotted, green lines mark the minimum, median, and maximum observed optical line ratios in the GP sample from \citet{cardamone09}. We have corrected these line ratios for Milky Way extinction using the \citet{cardelli89} law and for internal extinction using a \citet{calzetti00} law, as described in \citet{jaskot13}.}
\label{fig:gp_predict}
\end{figure*}

In Figure~\ref{fig:gp_predict}, we show the expected \ciii\ EWs of the GPs based on their observed \oiii~$\lambda$5007, \oii~$\lambda$3727, and \neiii~$\lambda$3869 line strengths and calculated metallicities (Z$\sim$0.003; \citealt{izotov11}). We correct the GPs' emission lines for Milky Way extinction using the \citet{cardelli89} law and for internal extinction using a \citet{calzetti00} law \citep[see][]{jaskot13}. 

The possible \ciii\ EWs for most of the GPs range from 2 to 10 \AA, depending on the age of the starburst. The most highly ionized objects have predicted EWs that are higher by about 1 \AA. Even for these high ionization parameters, however, the GPs' predicted \civ\ emission will likely reach only 50\%\ of the \ciii\ emission (see Figure~\ref{fig:c3ew_c4c2}). The models predict even weaker \heii~$\lambda$1640 emission, $< 10$\%\ of the \ciii. However, the models may not give accurate predictions for such high ionization species. For $Z=0.003$ and log $\bar{U}>-1$, the models predict \heii~$\lambda$4686/H$\beta$ ratios $\lesssim0.005$, whereas the GPs commonly have observed optical \heii~$\lambda$4686/H$\beta$ ratios of 0.01-0.02 \citep[\eg][]{hawley12, jaskot13}. The models therefore require additional sources of high energy photons, such as a larger Wolf-Rayet population, shocks, or high-mass X-ray binaries \citep[\cf][]{shirazi12}. 

Several diagnostics suggest that the GPs may have escaping LyC radiation \citep[\eg][]{jaskot13, jaskot14, nakajima14, verhamme15, henry15}. So far, five GPs have confirmed LyC detections, with measured escape fractions of 6-13\%\ \citep{izotov16, izotov16b}. Low optical depth in the GPs and similar galaxies could therefore alter their observed \ciii\ emission strengths. With \oiii/\oii\ ratios of $\sim1-10$ and $Z=0.003$, \ciii\ EWs below 1-4 \AA\ will imply low optical depths globally (Figure~\ref{fig:gp_predict}, lower panel), while \ciii\ EWs above 6 \AA\ would necessitate young ages of $<5$ Myr. We will test these predictions for the \ciii\ EWs with {\it Hubble Space Telescope (HST)} observations of the GPs using the Space Telescope Imaging Spectrograph ({\it HST} Program GO-14134, P.I.: Ravindranath). 

\section{Discussion}
\subsection{Implications of Existing \ciii\ Detections}

The fact that strong \ciii\ emission is common among low-metallicity galaxies attests to the importance of binary evolution effects. The high fraction of close binaries among massive stars \citep[\eg][]{sana12} and recent analyses of both low- and high-redshift galaxies \citep[\eg][]{wofford16, steidel16} also imply that interacting binaries may influence the emission from star-forming regions. We find that only the BPASS stellar population models reproduce the \ciii\ EWs $\geq 17$ \AA\ observed in some low-metallicity galaxies \citep{rigby15}. Furthermore, the binary models naturally explain why so many observed galaxies have \ciii\ EWs $\geq$10 \AA. In the Geneva and Padova single-star population models, such high EWs disappear only 2-3 Myr after the initial burst. These models would suggest that we are catching these high-redshift galaxies at a specific, unusual time in their evolution. In contrast, binary interactions can amplify galaxies' ionizing fluxes several Myr after a burst \citep[\eg][]{stanway16}, and the low-metallicity BPASS models can maintain EWs near 10\AA\ for 10 Myr after an instantaneous burst or 100 Myr of continuous star formation (Figure~\ref{fig:ew_bpass}). Binary interactions also provide an alternative channel for Wolf-Rayet star formation at low metallicity and may help produce the Wolf-Rayet features observed in several \ciii-emitters \citep{rigby15}. Binary evolution dramatically affects the ionizing output of low-metallicity galaxies, and models must take it into account to accurately determine high-redshift galaxy properties.

The high \ciii\ EWs commonly observed in high-redshift galaxies also imply high ionization parameters, log~$\bar{U}\geq-2$.  In contrast, most low-redshift galaxies have log~$U \leq-3$ \citep[\eg][]{lilly03, nakajima14}. The EWs therefore further support the notion that typical ionization parameters increase with redshift \citep[\eg][]{brinchmann08, richard11}.

\subsection{\ciii\ in $z>6$ Galaxies}

The photoionization models successfully explain the \ciii\ EWs observed in a variety of galaxies and shed light on the observability and possible uses of \ciii\ at $z>6$. In particular, the models demonstrate that strong \ciii\ emission should be common among high-redshift galaxies. The elevated temperatures at low metallicity enhance \ciii\ emission, even assuming that lower C/O ratios characterize low-metallicity systems. Binary interactions in the stellar population allow the \ciii\ emission to reach detectable levels for an extended period of time. While extreme ionization parameters will generate the strongest \ciii, even moderate ionization parameters of log~$\bar{U}=-3$ will produce \ciii\ EWs of at least a couple {\AA}ngstroms. Both observationally and theoretically, \ciii\ emission scales with other forbidden emission lines, including \oiii~$\lambda$5007. Many $z>6$ galaxies show evidence of enormous \oiii\ EWs ($\sim$600-1600 \AA; \citealt{labbe13}; \citealt{smit14}), and our models support the suggestion of \citet{stark14} that these galaxies will likely produce strong \ciii\ emission as well.

Although most $z=6-7$ galaxies should exhibit strong \ciii\ emission, certain physical conditions will not permit detectable \ciii. Galaxies with log~$\bar{U} \leq -4$ will produce weak \ciii, with EWs $<1$ \AA. Weakly star-forming galaxies or galaxies with extended gas distributions may therefore lack observable \ciii. Models with C/O$=0.04$ also typically produce \ciii\ EWs $\leq2$ \AA. Observationally, C/O ratios correlate with galaxy metallicity, likely because of a metallicity dependence in the stellar wind yields of massive stars \citep[\eg][]{akerman04, erb10}. At $Z\leq0.001$, our lowest modeled metallicity, observational constraints on galaxies' C/O ratios range from $\sim0.1-0.5$ \citep{garnett04, erb10, berg16}, which suggests that $Z=0.001$ galaxies should produce detectable \ciii. Notably, galaxies with $Z<0.001$, such as I Zw 18 and MACS0451-1.1 show \ciii\ EWs of 1.3-6.7 \AA\ (Figure~\ref{fig:ew_z_obs}), implying C/O $>0.04$. However, if C/O declines to values $\leq0.04$ at metallicities lower than those considered here, \ciii\ will likely cease to be easily observable.

Our analysis agrees with the findings by \citet{stark14} that \ciii\ offers the most reliable prospect for redshift identifications at $z>6$, where \lya\ weakens considerably. \ciii\ is almost always the strongest emission line at $<2700$ \AA\ or is comparable in strength to the strongest line. \ciii\ EWs reach levels of several \AA\ or more for a wide range of nebular conditions at low metallicities, and \ciii\ observations of most $z>6$ galaxies therefore stand a high chance of success. In addition, the \ciii-emitting population should provide a good proxy for LAEs, once the intergalactic medium opacity suppresses \lya. \citet{hagen16} find that galaxies with strong forbidden line emission are indistinguishable from LAE populations in most physical properties, and observationally, \lya\ EWs correlate with \ciii\ EWs \citep{stark14, rigby15}. The low dust content, high ionizing fluxes, and strong feedback of young, metal-poor starbursts may optimize \lya\ escape; these same physical conditions will also enhance the observed \ciii\ emission.

The \ciii~$\lambda$1909 doublet can serve as a key tool in studying the $z>6$ galaxies that likely reionized the universe. Bright star-forming galaxies at $z>6$, with $M_{\rm vir}\geq10^9$\Msol, may provide the dominant contribution to reionization at late times \citep[\eg][]{sharma16, xu16}. At $z=6$, these galaxies may be metal poor, but not completely devoid of metals, with 12+log(O/H)$\sim7.0-7.5$ \citep{ma16}. The cosmic reionizers may also exhibit particularly high ionization parameters. Six of the strongest confirmed LyC-leaking systems, with escape fractions \fesc$=6-50$\%, have extremely large \oiii/\oii\ ratios indicative of high ionization parameter \citep{izotov16b, vanzella16}. The young, low-metallicity, highly ionized galaxies that should show the strongest \ciii\ emission may therefore also be the galaxies responsible for reionization. However, as discussed in \S~\ref{sec:results:tau}, if \fesc\ is sufficiently high, it could weaken the observed \ciii\ emission. In extreme LyC-emitters, with \fesc\ $\gtrsim$50\%, \ciii\ may weaken below detectable levels. Current predictions place necessary \fesc\ values at $\lesssim20$\% \citep[\eg][]{robertson15, stanway16, ma16b, sharma16, xu16}, low enough to permit significant \ciii. The \ciii\ line itself may help identify LyC-leaking galaxies; objects with weaker than expected \ciii\ for their age, metallicity, and ionization parameters are good candidates for LyC escape.

\subsection{Prospects for {\it JWST}}
The James Webb Space Telescope ({\it JWST}) will play a major role in identifying and characterizing the galaxies in the reionization era. At $z>6$ the IGM opacity continues to increase, and as suggested by the drop in LAEs over $6<z<8$ \citep{pentericci14, tilvi14, konno14}, \lya\ emission may be severely quenched well into the reionization era. The \ciii~$\lambda$1909 doublet is a good alternative redshift indicator that is accessible for $z>3$ galaxies with {\it JWST} multi-object spectroscopy at $0.7-5.0\mu$m. 

In order to characterize the nature of the reionizers (star-forming galaxies or AGN), it will be important to rely on the emission-line diagnostics in the rest-frame UV-optical spectra that will be redshifted to the wavelengths accessible with {\it JWST}. At $z>8-9$, the optical emission lines (H$\beta$, \oiii~$\lambda\lambda$4959, 5007, H$\alpha$, and \nii~$\lambda\lambda$6548, 6583) will redshift beyond 5.0 $\mu$m, and only the rest-UV emission lines will be accessible to the wide-field spectroscopic modes on {\it JWST} (e.g., NIRSpec MSA, slitless spectroscopy with NIRISS, NIRCam). 

{\it JWST} has the sensitivity to easily detect multiple UV emission lines in large number of galaxies at $z>10$, such as \civ~$\lambda$1549, \oiiib~$\lambda\lambda$1661, 1666, and \ciii~$\lambda$1909, which are seen in emission in $z>6$ galaxies \citep{stark15, stark15b, stark16}. For $z>6$ galaxies, {\it JWST} will be able to resolve $\sim$ 200 pc at restframe UV ($\leq1500$\AA), and will easily resolve star-forming regions or clumps on sub-100 pc scales within strongly-lensed galaxies \citep[\eg][]{wuyts14}. Our photoionization models, which consider individual starburst regions, will offer direct interpretations for the emission-line spectra dominated by individual star-forming clumps or by multiple regions with similar properties. As discussed in \S~\ref{sec:results:linescalings}, the \civ/\ciii\ ratio can constrain the ionization parameter, and the C/O and \ciii\ EW will provide constraints on the age and metallicity (Figure~\ref{fig:c3ew_c4c2}). The UV emission-line ratios from photoionization models presented in this work (also see \citealt{gutkin16}) offer useful diagnostics to infer the ionization parameter, age and metallicity of the nebular regions and identify the role of shocks (Figure~\ref{fig:shockdiagnostic}). The suite of rest-UV emission lines also offer diagnostics for AGN at $z>10$ \citep{feltre16}. With {\it JWST}, these UV emission-line diagnostics will characterize the nebular conditions and ionizing sources in $z>10$ galaxies.

\section{Summary}
The \fciii~$\lambda$1907$+$\ciii~$\lambda$1909 doublet appears prominently in spectra of low-metallicity galaxies at both low and high redshift \citep[\eg][]{shapley03, stark14, rigby15}. We present CLOUDY photoionization model predictions for \ciii\ line strength, generate diagnostic grids using \ciii, and discuss the implications for observations of star-forming emission line galaxies. We summarize our results below:
\begin{enumerate}

\item We test predictions from three ionizing SEDs: Padova models for single, non-rotating stars, Geneva models for single stars including rotation, and BPASS models including binary interaction effects and rotation (Figures~\ref{fig:ew_pi}-\ref{fig:ew_bpass}). Only the BPASS models reproduce observed \ciii\ EWs above 17 \AA. The harder SEDs of the BPASS models also produce high \ciii\ EWs over longer timescales than the single-star models. Binary interaction effects are necessary to understand the prevalence of strong \ciii\ emission in low-metallicity galaxies.

\item The models predict that \ciii\ emission will peak at the youngest ages of a starburst. However, if dense natal gas obscures UV emission at early times, peak \ciii\ emission may arise slightly later, during the Wolf-Rayet phase, as implied by observations of low-redshift galaxies \citep{rigby15}.

\item Low metallicities and high ionization parameters enhance \ciii\ emission \citep[c.f.,][]{erb10, stark14, steidel16}. At fixed C/O ratio, \ciii\ EWs are strongest in the lowest metallicity models with $Z=0.001$. The models support the existence of a metallicity threshold of $Z\sim$0.006 for \ciii\ emission \citep{rigby15}; above this threshold, \ciii\ EWs do not exceed 5 \AA\ (Figure~\ref{fig:ew_z_obs}). 

\item Starburst age, metallicity, and ionization parameter are the dominant causes of strong \ciii\ emission; in the low metallicity models ($Z\lesssim0.004$), lower dust content, higher nebular densities, and shell-like geometries also enhance \ciii, but to a lesser extent. Shock ionization, predicted from the Mappings III models \citep{allen08}, may also raise \ciii\ fluxes, particularly for shocks with lower velocities or higher magnetic field strengths.

\item Optically thin, density-bounded models exhibit weaker \ciii\ fluxes, with high ionization parameter models showing the steepest declines in \ciii\ flux with optical depth. For LyC escape fractions of 20\%, \ciii\ fluxes are weaker by 2-68\%. Weaker-than-expected \ciii\ EWs for a given ionization parameter could help select strong LyC-leaking systems.

\item \ciii\ EWs correlate only weakly with the intrinsic \lya\ EWs produced from the \hii\ region. However, the same conditions responsible for strong \ciii, such as high ionization parameter or low metallicity and a correspondingly low dust content, may promote \lya\ escape.

\item Both the photoionization models and observed galaxies indicate a correlation between \ciii\ emission and \oiii~$\lambda$5007. The photoionization models predict the strongest correlation between \ciii\ and the temperature-sensitive \oiiib~$\lambda\lambda$ 1661,1666 and \oiii~$\lambda$4363 lines, which suggests that the direct abundance method could prove more feasible in \ciii-emitting galaxies at high redshift. With constraints on C/O from \ciii\ and \oiiib~$\lambda$1665, the \ciii~$\lambda$1909/\oiii~$\lambda$4363 could constrain dust attenuation.

\item Nebular \heii~$\lambda$1640 emission is weak in the models, with a maximum EW $< 1$ \AA\ for pure photoionized gas. However, \heii\ is sensitive to even low levels of shock emission. We show that the \ciii/\heii\ vs.\ \civ~$\lambda$1549/\heii\ diagnostic proposed for AGN \citep{villarmartin97, allen98, feltre16} also separates star-forming regions from shock-ionized gas (Figure~\ref{fig:shockdiagnostic}).

\item \ciii\ is the strongest line at $<2700$ \AA\ after \lya\ in most of the models explored in this paper, which corroborates the idea that \ciii~$\lambda$1909 represents the best option for spectroscopic redshift identification at $z>6$ \citep{stark14}. Furthermore, \ciii\ reaches detectable levels ($> 2$ \AA) for a wide range of conditions at low metallicity. 

\item Extreme emission line galaxies with high ionization parameters should emit strong \ciii. Recent observations of highly ionized starburst galaxies at low redshift suggest elevated LyC escape fractions in these systems \citep{izotov16, izotov16b}. If galaxies at $z>6$ behave in a similar manner, these results suggest that the galaxies responsible for reionization will have detectable \ciii\ emission for moderate LyC escape fractions (\fesc $\lesssim30-40$\%). 

\end{enumerate}

\acknowledgments{We thank Harry Ferguson, Jane Rigby, Jason Tumlinson, and Aida Wofford for interesting and helpful discussions. A.E.J.\ acknowledges support from NASA through grant HST-GO-14134 from STScI, which is operated by AURA under NASA contract NAS-5-26555. }

\appendix
\section{Effects of Additional Nebular Parameters}
\label{appendix:neb}
\subsection{C/O Ratio}
\label{appendix:co}
\begin{figure*}
\plotone{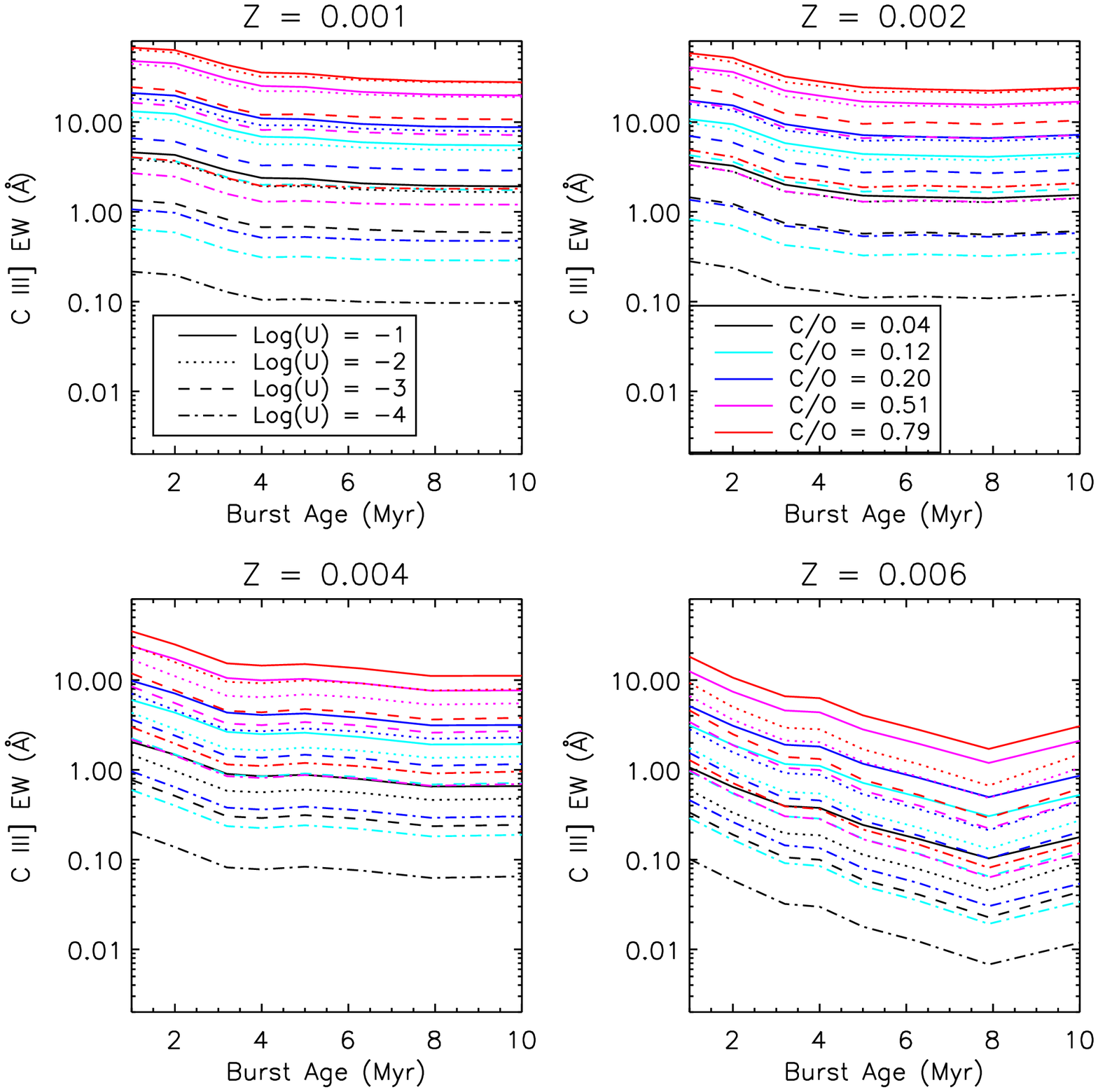}
\caption{Predicted \ciii\ EWs from the BPASS models as a function of instantaneous burst age, metallicity, $\bar{U}$, and C/O abundance ratio. Each panel corresponds to a different metallicity. Line style denotes $\bar{U}$, and color indicates the C/O ratio. Other parameters are as specified in Table~\ref{table:fiducial}. }
\label{fig:ew_co}
\end{figure*}

In Figure~\ref{fig:ew_co}, we show the predicted \ciii\ strengths for models with different C/O ratios. Interestingly, \ciii\ emission does not quite increase linearly with C/O; for instance, a five-fold decrease in C/O from 0.20 to 0.04 at $Z=0.001-0.002$ and log $\bar{U} \geq -2$ decreases the \ciii\ EW by a factor of 4.5-4.8. Since carbon is a coolant, the lower carbon abundance raises the electron temperature, and the increased collisional excitation rate of \ciii\ partially compensates for the reduced C abundance. 

Observations of low-metallicity dwarf galaxies by \citet{berg16} suggest that C/O ratios decline with decreasing metallicity but stabilize at a constant value of $\sim$0.2 below $Z\sim0.003$. However, individual galaxies may show significant variation in their C/O ratios \citep{berg16}. High-redshift galaxies have C/O ratios in the range of $\sim0.15-0.28$ \citep{shapley03}, similar to the range observed in local dwarf irregular galaxies \citep{garnett95, garnett97, kobulnicky98, berg16}. Nevertheless, the relationship between $Z$ and C/O shows substantial scatter, and individual galaxies may exhibit higher or lower C/O ratios than expected. If we adopt a declining C/O ratio with metallicity such that C/O drops from 0.25 to 0.15 by Z=0.001 \citep{erb10}, then peak \ciii\ emission should occur at $Z=0.002$, rather than $Z=0.001$. 

\subsection{Dust Content}
\label{appendix:dust}

\begin{figure*}
\plotone{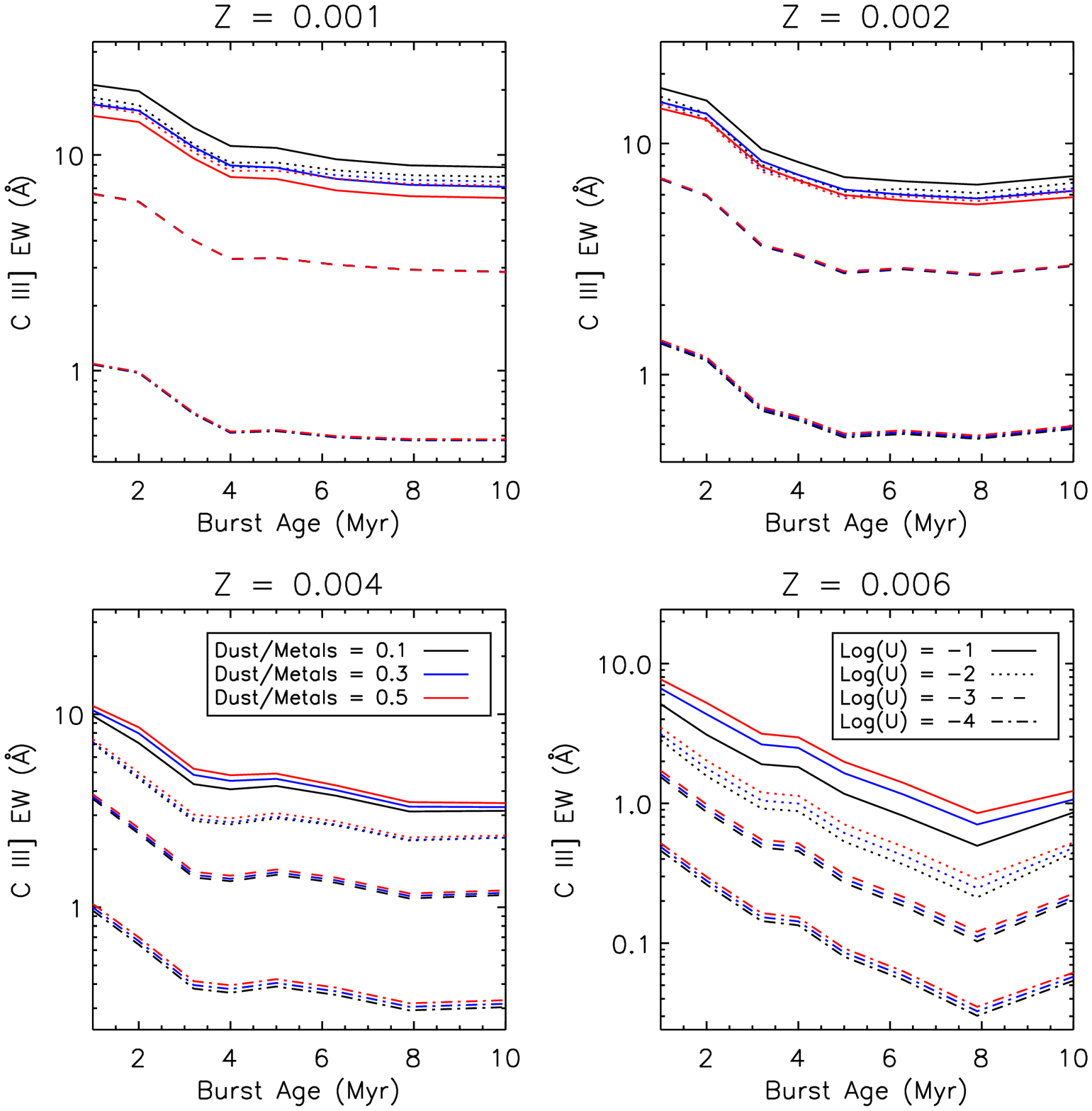}
\caption{Predicted \ciii\ EWs from the BPASS models as a function of instantaneous burst age, metallicity, $\bar{U}$, and dust-to-metals ratio. Each panel corresponds to a different metallicity. Line style denotes $\bar{U}$, and color indicates the dust-to-metals ratio.  Other parameters are as specified in Table~\ref{table:fiducial}. }
\label{fig:ew_dust}
\end{figure*}

In Figure~\ref{fig:ew_dust}, we show the predicted emergent \ciii\ EWs for dust-to-metals ratios of 0.1, 0.3, and 0.5. Increasing the dust-to-metals ratio raises the electron temperature of the gas. The higher dust abundance increases the photoelectric heating rate and lowers the cooling from forbidden lines by depleting metals onto dust grains.  Although the higher temperature gas could raise the \ciii\ emission, for $Z\leq0.003$, dust extinction ultimately dominates; the models with higher dust show weaker emergent \ciii\ EWs. At higher metallicities, $Z\geq0.004$, however, dust heating proves more important than dust extinction, and the models with higher dust content show higher emergent \ciii\ EWs. We caution that the actual degree of \ciii\ suppression due to extinction will depend on the details of the dust geometry and reddening law, which we do not consider here.

\subsection{Nebular Geometry}
\label{appendix:geometry}

\begin{figure*}
\plotone{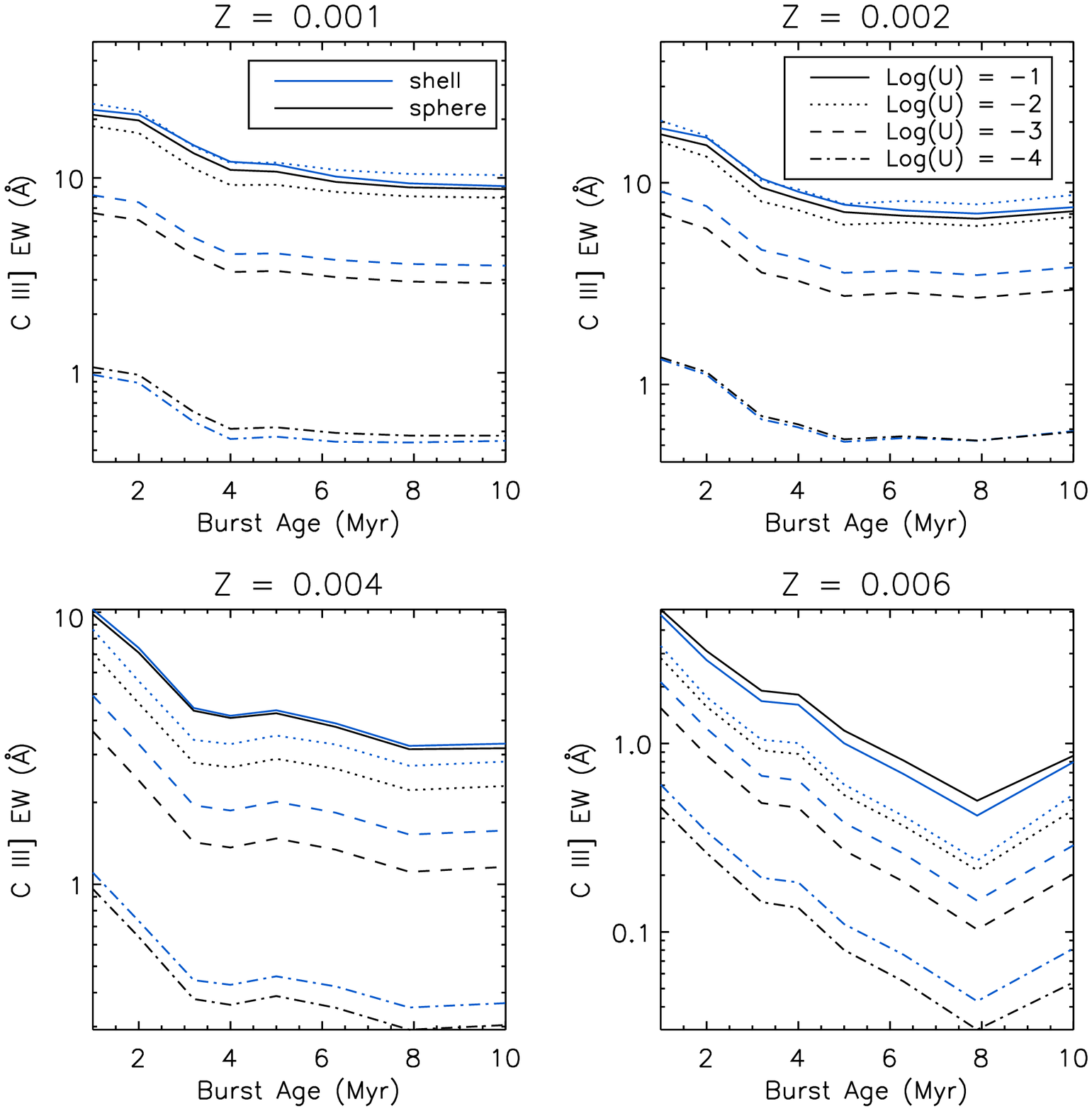}
\caption{Predicted \ciii\ EWs from the BPASS models as a function of instantaneous burst age, metallicity, $\bar{U}$, and geometry. Each panel corresponds to a different metallicity. Line style denotes $\bar{U}$, blue lines correspond to a shell geometry, and black lines correspond to a filled sphere geometry. Other parameters are as specified in Table~\ref{table:fiducial}. }
\label{fig:ew_geom}
\end{figure*}

The adopted nebular geometry weakly affects the \ciii\ emission. Figure~\ref{fig:ew_geom} compares the predictions of the filled sphere and hollow shell geometries. The shell model EWs range from $\sim$0.9 to $\sim$1.6 times the predictions of the filled sphere models, with the shell nebulae typically showing stronger \ciii\ emission. In particular, all models with high ionization parameter (log $\bar{U} \geq -3$) and low metallicity ($Z \leq 0.003$) show higher \ciii\ emission for the shell geometry.

\end{document}